# Model composition through model reduction: a combined model of CD95 and NF-κB signaling pathways

**Elena Kutumova[1,2*], Andrei Zinovyev[3,4,5], Ruslan Sharipov[1,6] and Fedor Kolpakov[1,2]**

[1]Institute of Systems Biology, Ltd, 15 Detskiy proezd, Novosibirsk 630090, Russia

[2]Design Technological Institute of Digital Techniques, The Siberian Branch of The Russian Academy of Sciences, 6 Acad. Rzhanov Str., Novosibirsk 630090, Russia

[3]Institute Curie, 26 rue d'Ulm, Paris 75248, France

[3]Institut Curie, 26 rue d’Ulm, F-75248 Paris France

[4]INSERM, U900, Paris, F-75248 France

[5]Mines ParisTech, Fontainebleau, F-77300 France

[6]Institute of Cytology and Genetics, The Siberian Branch of The Russian Academy of Sciences, 10 Acad. Lavrentyev Ave., Novosibirsk 630090, Russia

[*]Corresponding author

Email addresses:

EK: elena.kutumova@biouml.org

AZ: andrei.zinovyev@curie.fr

RS: shrus79@biouml.org

FK: fedor@biouml.org

# Abstract

**Background.** Many mathematical models characterizing mechanisms of cell fate decisions have been constructed recently. Their further study may be impossible without development of methods of model composition, which is complicated by the fact that several models describing the same processes could use different reaction chains or incomparable sets of parameters. Detailed models not supported by sufficient volume of experimental data suffer from non-unique choice of parameter values, non-reproducible results, and difficulty of analysis. Thus, it is necessary to reduce existing models to identify key elements determining their dynamics, and it is also required to design the methods allowing us to combine them.

**Results.** Here we propose a new approach to model composition, based on reducing several models to the same level of complexity and subsequent combining them together. Firstly, we suggest a set of model reduction tools that can be systematically applied to a given model. Secondly, we suggest a notion of a minimal complexity model. This model is the simplest one that can be obtained from the original model using these tools and still able to approximate experimental data. Thirdly, we propose a strategy for composing the reduced models together. Connection with the detailed model is preserved, which can be advantageous in some applications. A toolbox for model reduction and composition has been implemented as part of the BioUML software and tested on the example of integrating two previously published models of the CD95 (APO-1/Fas) signaling pathways. We show that the reduced models lead to the same dynamical behavior of observable species and the same predictions as in the precursor models. The composite model is able to recapitulate several experimental datasets which were used by the authors of the original models to calibrate them separately, but also has new dynamical properties.

**Conclusion.** Model complexity should be comparable to the complexity of the data used to train the model. Systematic application of model reduction methods allows implementing this modeling principle and finding models of minimal complexity compatible with the data. Combining such models is much easier than of precursor models and leads to new model properties and predictions.

## Background

Systems biology aims to study complex interactions in living systems and focuses on analysis and modeling their properties. Mathematical modeling provides several ways to describe biological processes based on experimental information of different kind. However, creation of detailed models not supported by enough experimental data often makes their analysis and interpretation difficult [1]. Several aspects of the same process can be modeled using different levels of abstraction involving different reaction chains, chemical kinetics, and incomparable sets of parameters. Such models are difficult to merge. Meanwhile, merging is an important approach for creation of complex models. Thus, development of efficient methods and software, allowing us to combine models, is the object of intense study in systems biology. In our work, we focused on the fact that, generally, complexity of models is not comparable to the volume of experimental data used to adjust their parameters. Due to this fact, we turn to the methods of model reduction allowing us to minimize model's complexity without affecting the model simulation dynamics.

Model reduction is a well-established technique in many fields of biochemical research and engineering. It has been used for many years in chemical kinetics (for reviews, see [2-4]) and has already found multiple applications in systems biology, including discrete modeling [5] and modeling of metabolic pathways [6, 7]. The principles of this technique have been applied in computational biology [8] and implemented as a part of widely used pathway simulators such as BioUML [9], BIOCHAM [10], COPASI [11], and GINsim [12]. Model reduction led to new insights in mechanisms of translation regulation by microRNAs [13, 14] and was applied for analysis of such signaling pathways as JAK-STAT [15], NF-κB [16], and EGFR [17].

In our investigation, we used the principles of model reduction to construct reasonably accurate minimal size approximations of two different models describing the CD95 signaling pathways [18, 19]. The first model explores pro-apoptotic properties of CD95 after stimulation by its natural ligand CD95L or by agonistic antibodies anti-CD95 implying formation of the death-inducing signaling complex (DISC) [18, 20]. DISC consists of oligomerized CD95, death domain-containing adaptor FAS-associated molecule (FADD), procaspases-8 and -10, and two isoforms of cellular FLIP (CFLAR) protein (cFLIP long and cFLIP short). Caspase-8 leads to activation of effector caspase-3 directly (type I cells) or via stimulation of cytochrome *C* release from the mitochondria (type II cells) [21]. The latter step requires formation of the apoptosome complex and activation of caspase-9. Once activated,

caspase-3 cleaves poly(ADP-ribose) polymerase (PARP), thereby making the apoptotic process irreversible. The second model describes the state when CD95 not only activates the pro-apoptotic pathway, but also induces transcription factor NF-κB that is an important regulator of cell survival functioning [19]. This is possible due to cFLIPL cleavage in the DISC complex. The cleaved p43-FLIP directly interacts with the IKK complex and activates it. The activated IKK performs phosphorylation of the IκB protein and thereby frees NF-κB.

The authors of the models have evaluated concentration changes of major apoptotic molecules by Western blot analysis and represented them as relative values at given time points. Using the systematic methodology [2] implemented in the BioUML software, we reduced the models so that they still satisfy these data. This allowed us to simplify the overlapping components of the models and find a way for their composition.

# Methods

## Model reduction

Mathematical modeling of biological processes based on the classical theory of chemical kinetics assumes that a model consists of a set of species $S = (S_1, \dots, S_m)$ associated with a set of variables $C(t) = \big(C_1(t), \dots, C_m(t)\big)$ depending on time $t \in [0, T]$, $T \in R^+$, and representing their concentrations, and a set of biochemical reactions with rates $v(t) = \big(v_1(t), \dots, v_n(t)\big)$ depending on a set of kinetic constants $K$. Reaction rates are modeled by mass-action or Michaelis-Menten kinetics. A system of ordinary differential equations (ODE) representing a linear combination of reaction rates is used to describe the model behavior over time:

$$\frac{dC(t)}{dt} = N \cdot v(C, K, t), \qquad C(0) = C^0. \tag{1}$$

Here $N$ is a stoichiometric matrix of $n$ by $m$. We say that $C^{ss}$ is a steady state of the system (1) if

$$N \cdot v(C^{ss}, K, t) = 0, \qquad \lim_{t \to \infty} C_i(t) = C_i^{ss}. \tag{2}$$

Model reduction implies transformation of the ODE system to another one with smaller number of equations without affecting dynamics of variables $C_1(t), \dots, C_s(t)$, which is fixed by a set of experimental points $C_i^{exp}\big(t_{ij}\big)$ at given times $t_{ij}$, $j = 1, \dots, r_i$, where $r_i$ is number of such points for the concentration $C_i(t)$, $i = 1, \dots, s$. To check dynamics preservation in the course of model reduction, we consider the function of deviations defined as a normalized sum of squared differences [11]:

$$f_{dist}(C^0, K) = \sum_{i=1}^{s}\sum_{j=1}^{r_i}\frac{\omega_{min}}{\omega_i}\cdot\left(C_i(t_{ij}) - C_i^{exp}(t_{ij})\right)^2, \tag{3}$$

where $\omega_{min} = \min_i \omega_i$, weights $\omega_i = \sqrt{r_i^{-1}\cdot\sum_j\left(C_i^{exp}(t_{ij})\right)^2}$ are calculated as mean squared values of experimental concentrations for each variable and the normalizing factor $\omega_{min}/\omega_i$ is used to make all concentration trajectories to have similar importance.

Reducing kinetic model is possible when some quantities are much smaller than other quantities and can be neglected. Usually, this implies some qualitative relations (much bigger, much smaller) between model parameters. When these relations satisfy certain rules, we can approximate the detailed model by a simpler one. If the parameters are (approximately) determined, as in our case, we find an approximation specific for a region of the parameter space. However, if we want to investigate the model behavior for the entire space, we need to decompose it into regions characterized by asymptotically different behaviors of the dynamical systems. Afterwards, a specific reduction should be performed for each region (fig. 1).

We define the *minimal complexity model* as a reduced model with the minimal number of elements (species and reactions) so that the deviation function value (3) calculated for the reduced model is different from the original model no more than 20%. Such threshold is explained by the fact that in this work we considered experimental data obtained by Bentele *et al.* [18] and Neumann *et al.* [19] using Western blot technology with the standard deviation 15-20%.

Reduction of mathematical model complexity is achievable by different methods [2]. Description of the methods used in our work is provided below.

**(MR1)** *Removal of slow reactions*. In the simplest case, one can remove a reaction which rate does not change significantly the amount of any species, during the duration $T$ of the computational simulation (i.e., its rate $v$ normalized on the characteristic amount of any reactant or product involved in it is much less than $1/T$). In more complex case, sometimes it is possible to neglect slow reactions whose effect is dominated by faster reactions. We say that the reaction $r_1$ with rate $v_{r_1}(t)$ is much slower than the reaction $r_2$ with rate $v_{r_2}(t)$ if $\max_t|v_{r_1}(t)| < k\cdot\max_t|v_{r_2}(t)|$ during the simulation period $T$ ($0 \le t \le T$), where $k = 10^{-2}$. If a chemical species is a reactant in two reactions $r_1$ and $r_2$, and $r_1$ is much slower than $r_2$, then one can test the effect of removal of $r_1$, and eliminate it in some cases, if its removal does not significantly change the model dynamics. Note that it is not always possible due to existence of slowly relaxing cyclic subsystems of faster reactions, see [2]. Also note that in some cases, it is sufficient to consider $k = 10^{-1}$ or even $k = 1$.

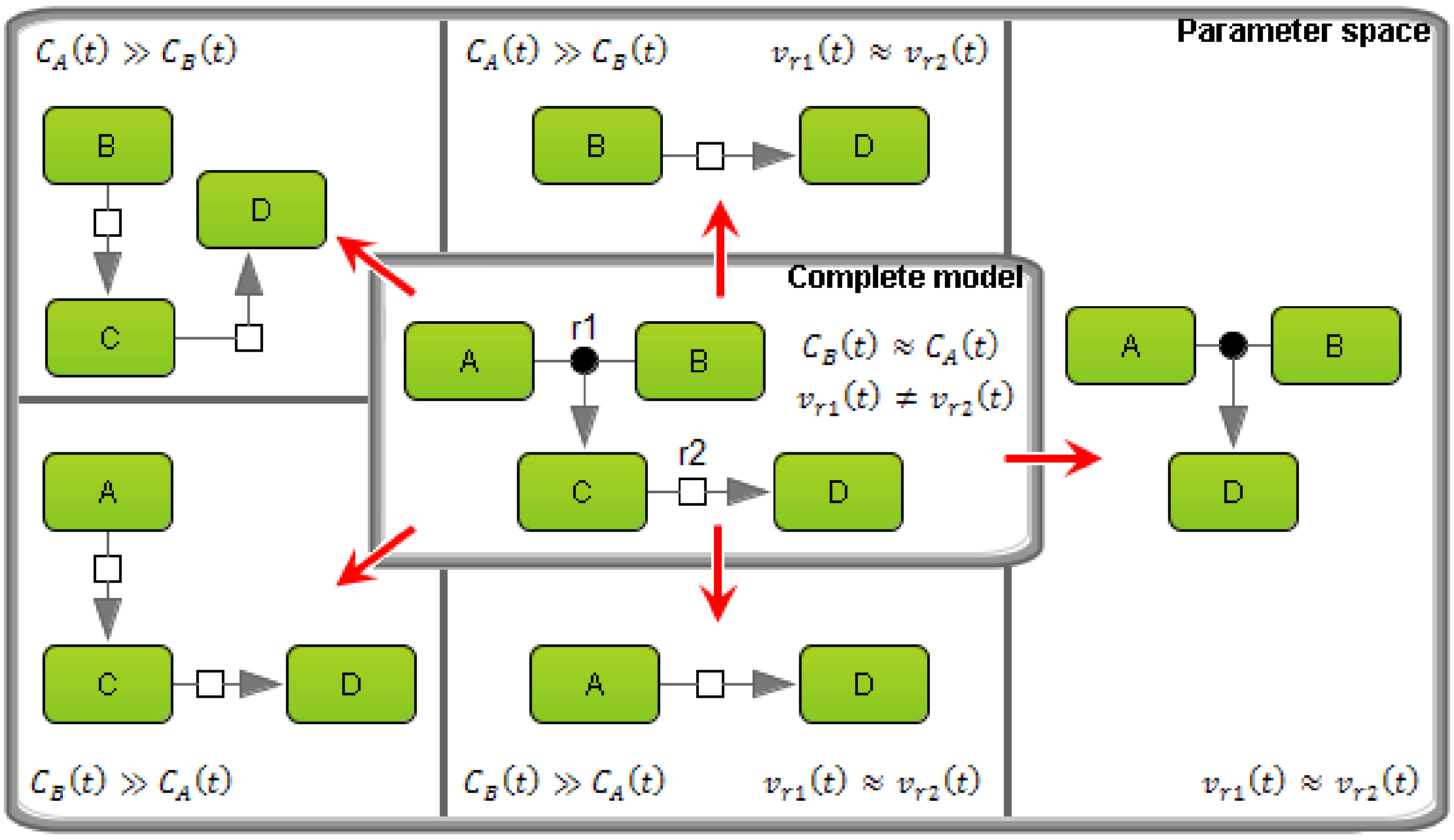

**Figure 1 – Schematic representation of the model reduction techniques**

A model consisting of four species ($A$, $B$, $C$, and $D$) and two reactions ($r1$ and $r2$) was reduced taking into account the relations between time-dependent concentrations $C_A(t)$ and $C_B(t)$, as well as reaction rates $v_{r1}(t)$ and $v_{r2}(t)$. In the case when $C_A(t) \gg C_B(t)$, the concentration of $A$ can be considered constant for some period of time. The same simplification could be applied to B.

**(MR2)** *Quasi-steady-state (QSS) approximation* [2, 16] assumes that the variables $C(t)$ of the model can be split in two groups: "slow" basic components $C^s(t)$ and fast intermediates $C^f(t)$. The amounts of fast intermediates are assumed to be smaller than the amounts of basic components, while the reaction rates are of the same order, or both fast intermediates and slow components participate in the same reactions. In this situation, one can rescale the intermediate amounts, introducing new variables $\tilde{C}^f(t) = \frac{1}{\varepsilon} C^f(t)$, where $\varepsilon$ is a small number such that $\tilde{C}^f$ and $C^s$ are of the same order now. For the new variables one can write

$$\frac{dC^s(t)}{dt} = W^s\left(C^s, \tilde{C}^f, K\right), \quad \frac{d\tilde{C}^f(t)}{dt} = \frac{1}{\varepsilon} \cdot W^f\left(C^s, \tilde{C}^f, K\right),$$

where $W^s$ and $W^f$ are functions that can be obtained from (1), having values and derivatives of the same order of magnitude during the simulation period and for the given set of parameters $K$. If this is the case then in the limit $\varepsilon \to 0$ one can substitute differential equations for $\tilde{C}^f$ by algebraic equations $W^f\left(C^s, \tilde{C}^f, K, t\right) = 0$, find dependencies $\tilde{C}^f =$ $\tilde{C}^f(C^s, K)$ and exclude $\tilde{C}^f$ from the remaining model equations. Classical example of application of QSS is the derivation of Michaelis-Menten formula for enzymatic catalysis reaction $S + E \leftrightarrow S{:}E \rightarrow P + E$, for the case when the amount of enzyme $E$ is much smaller than the amount of the substrate $S$, while both $S$ and $E$ participate in the same reaction of complex $S{:}E$ formation [22].

**(MR3)** *Lumping analysis* refers to reducing the number of model variables by grouping some species. In particular, this can be illustrated by reactions

$$r1{:}\, A + S_1 \rightarrow P_1, \qquad r2{:}\, A + S_2 \rightarrow P_2 \tag{4}$$

with kinetic rates $v_{r_1}(t) = k_1 \cdot C_A(t) \cdot C_{S_1}(t)$ and $v_{r_2}(t) = k_2 \cdot C_A(t) \cdot C_{S_2}(t)$. If $k_1 = k_2$ and $C_{S_1}(t) = C_{S_2}(t)$, then the system (4) can be replaced by a single reaction

$$2 \cdot A + S \rightarrow P,$$

where $C_S(t) = C_{S_1}(t) = C_{S_2}(t)$, $C_P(t) = C_{P_1}(t) = C_{P_2}(t)$ and $k = k_1 = k_2$. Note, that in this example lumped variables $C_{S_1}(t)$, $C_{S_2}(t)$ and $C_{P_1}(t)$, $C_{P_2}(t)$ are linearly dependent: however, this is not required in general case. For conditions on lumpability in monomolecular reaction networks, see [23].

**(MR4)** *Removal of approximately linearly dependent variables s*. If variables $A(t)$ and $B(t)$ (species concentrations or reaction rates of the model) are approximately linearly dependent:

$$A(t) \approx k \cdot B(t), \qquad t \in [0, T], \qquad k = const,$$

then we can replace one of them by another in all kinetic laws of the model. Note that this can be considered as a particular case of lumping.

**(MR5)** *Simplification of the Michaelis-Menten kinetics*. Consider a reaction $r$ of the form $S$ –$E \rightarrow P$, where an enzyme $E$ converts a substrate $S$ into product $P$. Reaction rate of such reaction is frequently defined by the Michaelis-Menten formula

$$v_r(t) = \frac{k \cdot C_S(t) \cdot C_E(t)}{Km + C_S(t)} \tag{5}$$

where the enzyme concentration is a dynamic variable allowing to use the same kinetics in different regions of the phase space. If $Km \gg C_S(t)$, then we can reduce this formula to the form: $v_r(t) = \frac{k \cdot C_S(t) \cdot C_E(t)}{Km}$. On the other hand, if $C_S(t) \gg Km$, then (5) can be replaced by the equation $v_r(t) = k \cdot C_E(t)$.

**(MR6)** *Simplification of equations based on the law of mass action when one reactant dominates others*. Consider a reaction $r$ of the form $S_1 + S_2 \rightarrow P$ with the kinetic law

$$v_r(t) = k \cdot C_{S_1}(t) \cdot C_{S_2}(t). \quad \textbf{(6)}$$

When $C_{S_1}(t) \gg C_{S_2}(t)$ for $t \in [0, T]$, the formula (6) can be replaced by the linear equation $v_r(t) = k \cdot C_{S_1}(0) \cdot C_{S_2}(t)$. The time $T$ of validity of such pseudo-lineary approximation is defined as a period during which the relative change of $C_{S_1}(t)$ does not exceed some $\varepsilon$, i.e. $k \int_0^T C_{S_2}(t) dt < \varepsilon$.

One way to perform the model reduction is to apply the foregoing methods in the numerical order (fig.2). Note that the methods MR5 and MR6 are of the same type, so it does not matter which one to use first. This way assumes that we calculate the value of the distance function (3) at each reduction step to test whether approximation of the experimental data is still within allowable limit or not.

## Model analysis and comparison

*Model comparison using Akaike information criterion ($AIC$).* This criterion [24] defines the relative complexity of a model based on the goodness of experimental data fit and the number of model parameters $|K|$ (initial concentrations of species are not considered):

$$AIC = \chi^2 + 2 \cdot |K|. \quad \textbf{(7)}$$

The function $\chi^2$ is defined by the formula (3) with weights $1/\sigma_{ij}^2$ (instead of $\omega_{min}/\omega_i$) [15], where mean deviations $\sigma_{ij}$ of experimental values $C_i^{exp}(t_{ij})$ are calculated using smoothing spline [25]:

$$\sigma_{ij}^2 = \frac{1}{4} \sum_{k=-2}^{2} \left( C_i(t_{i,j+k}) - C_i^{exp}(t_{i,j+k}) \right)^2.$$

For $j + k < 0$ and $j + k > r_i$ we assumed $C_i(t_{i,j+k}) = C_i^{exp}(t_{i,j+k}) = 0$.

Models can be compared using the Akaike criterion, if they approximate the same experimental data. When the $AIC$ value of one model is less than of others, we say that this model is simpler in terms of this criterion.

*Model comparison with the mean $AIC$.* When we want to compare models approximating different sets of experimental data, we could calculate the mean $AIC$ coefficients by the formula

$$AIC_{mean} = \frac{AIC}{n_{exp}} \quad \textbf{(8)}$$

where $n_{exp}$ determines the number of experimental points.

*Steady-state analysis* finds values of species concentrations according to the rules (2). It is desirable to reduce a model so that steady-state concentrations of all experimentally measured chemical species have not changed significantly.

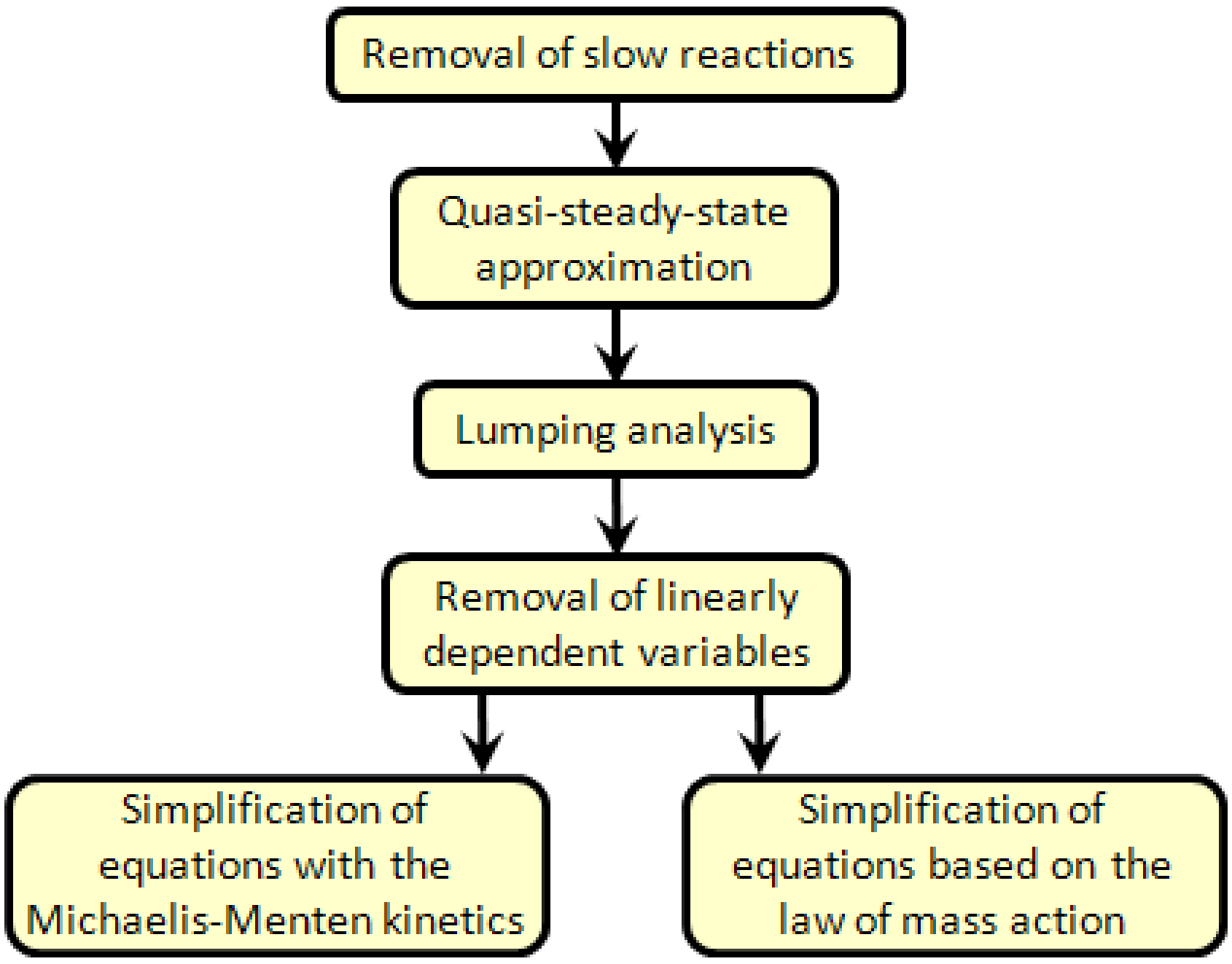

**Figure 2 – Flow chart of the model reduction**
The methods of model reduction are presented in the order of their application. The last two methods are of the same type, so this is a choice of the systems biologist which one to use first.

*Sensitivity analysis* reveals steady-state concentrations response to parameter perturbations. The local sensitivities $C_i^{ss}$, $i = 1, \dots, m$, for all parameters $p_j$, $j = 1, \dots, |K|$, are calculated via finite difference approximations:

$$S_{ij} = \frac{\partial C_i^{ss}}{\partial p_j} = \frac{C_i^{ss}(p_j + \Delta p_j) - C_i^{ss}(p_j)}{\Delta p_j}$$

It is useful to compute the mean log sensitivity for all fitted species:

$$S^{mean} = \frac{\sum_{i,j} \ln|S_{ij}|}{n_p \cdot s}, \qquad S_{ij} \neq 0, \qquad i = 1, \dots, s, \qquad j = 1, \dots, n_p, \qquad \textbf{(9)}$$

where $n_p$ is the number of all parameters $|K|$ or the number of parameters retained after the model reduction.

### Modular modeling

Considering the mathematical models of CD95 signaling pathways [18, 19], we decomposed them into functional modules. This step allowed us to identify overlapping components of the models and simplify their analysis. We defined a module as a submodel (including several species, reactions and parameters of the model) with input, output and contact ports. The first two types of ports characterized variables calculated in one module and passed to another through a directed connection. The contact ports declared common variables of modules via undirected connections (for more details of modular modeling, see [26]).

## Results

### Preliminary analysis, problems and inconsistencies in the precursor models

The mathematical model by Bentele *et al.* (fig. 3A) consists of 43 species including a virtual variable $x_{aa}$ intended for quantification of "apoptotic activity" caused by active caspases-3, -6 and -7 and calculated by the formula

$$\frac{dx_{aa}}{dt} = \frac{(1 - x_{aa}) \cdot \left(k_{36act} \cdot C_{casp3} + k_{36act} \cdot C_{casp6} + k_{7act} \cdot C_{casp7}\right)}{Km_{367act} + (1 - x_{aa})} - k_{dd} \cdot x_{aa}, \quad \textbf{(10)}$$

where $k_{36act}, k_{7act}, Km_{367act} \in K$. The variable $x_{aa}$ specifies a process of degradation in the model introduced as an exponential decay function $f_{degr}(x_{aa})$. This function is defined by the formula $k_d \cdot x_{aa}^2 + k_{ds}$ with $k_d$, $k_{ds} \in K$ for all active caspases and complexes containing their cleaved products, and by the formula $k_d \cdot x_{aa}^2$ for all other molecules besides cPARP (cleaved PARP), where $f_{degr}(x_{aa})$ is constant. All species in the model are degraded with the exception of cytochrome *C* and Smac stored in the mitochondria, whose concentrations are constant.

In total, the model contains 80 reactions (Table 1) including 24 reactions based on mass action kinetics, 12 reactions taken with kinetics of Michaelis-Menten, 41 reactions of degradation, the reaction of $x_{aa}$ production specified above and two reactions of cytochrome *C* and Smac release from the mitochondria modeled using a discrete event. The latter implies complete cytochrome *C* and Smac release within 7 minutes as soon as tBid reaches a certain level in comparison to Bcl-2/Bcl-XL. Since the authors did not provide exact form of the release function, we proposed the sigmoid function which has the expected behavior:

$$f_{release}(t) = 1 - \frac{1}{1 + \exp\left(k_{contr} \cdot \left(-t + t_{trigger} + 0.5 \cdot t_{release}\right)\right)} \quad \textbf{(11)}$$

where $t_{trigger}$ is the time when tBid concentration reaches a value of Bcl-2/Bcl-XL, $t_{release}$ is the start time of release and $k_{contr}$ is the contraction coefficient.

The model comprises 43 species (including $x_{aa}$) and 45 kinetic parameters, which the authors estimated based on the experimental data obtained by Western blot analysis for the human cell line SKW 6.4. Cells were stimulated by 5 μg/ml and 200 ng/ml of anti-CD95 (fast and reduced activation scenarios, respectively) and dynamics of several proteins (Bid, tBid, PARP, cPARP, procaspases-2, -3, -7, -8, -9, cleaved product of procaspase-8 p43/p41, and caspases-8) were measured.

We could not reproduce the dynamics of the original model using the parameter values provided by the authors. In particular, there was too rapid consumption of procaspases-2, -3, -7, -8 in the case of 5 μg/ml of anti-CD95 and procaspases-2, -9 in the case of 200 ng/ml. Degradation rates of procaspase-8 and caspases-8 was insufficient for both activation scenarios. Thus, we had to make several modifications of the original model to obtain the same dynamics as described in the original paper. Namely, we multiplied the rate constants of all the bimolecular reactions by the value $5 \cdot 10^{-5}$ and specified $k_d$ equal to 3.56 $min^{-1}$ and 0.62 $min^{-1}$ (instead of 0.891 $min^{-1}$ and 0.184 $min^{-1}$) for fast and reduced activation scenarios, respectively.

The model by Neumann *et al*. includes 23 species, 23 reactions constructed according to the law of mass action, and 17 kinetic parameters (fig. 4A). It reproduces experimental measurements of 8 entities (total amount of procaspase-8, its cleaved product p43/p41, caspase-8, procaspase-3, caspase-3, IκB-α, phosphorylated IκB-α and cleaved p43-FLIP) obtained using Western blot technology for HeLa cells stably overexpressing CD95–GFP and treated with three different concentrations of agonistic anti-CD95 antibodies (1500, 500 and 250 ng/ml). This is the reduced model suggested by the authors for description of the CD95-mediated pathways of cell death and NF-κB activation. Model reduction was based on intuition about what is important in the biochemical mechanisms. It was performed as long as the parameter estimation procedure (repeated after each step of the reduction) was able to provide a good fit of the experimental data. Particularly, the details of some complex formations were omitted as well as some non-measured species were removed. The authors submitted the model to the BioModels database [27]. Thus, we had no problems reproducing it.

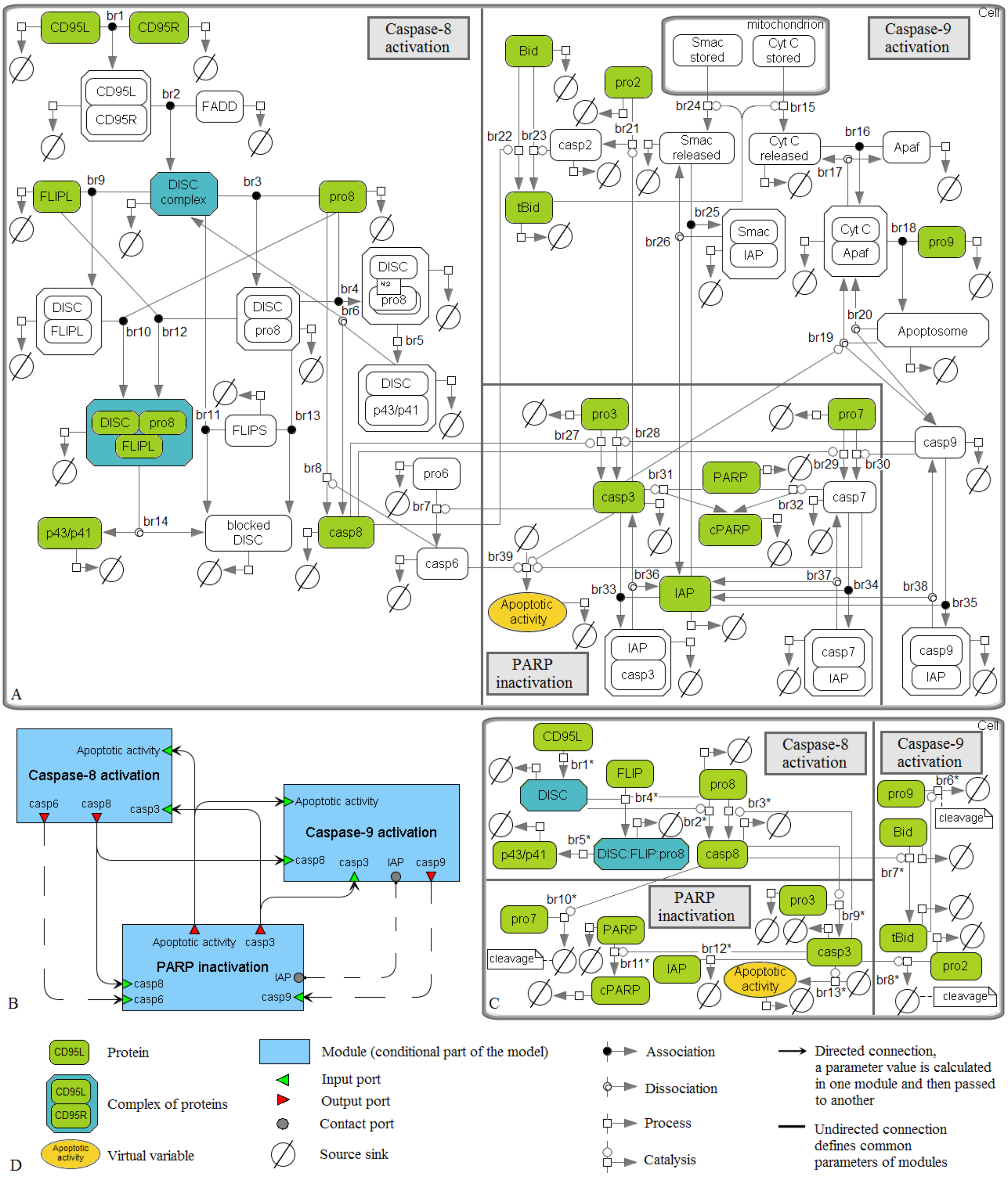

**Figure 3 – The model by Bentele *et al*. and results of its reduction**

**A.** The original model decomposed into modules according to three steps of apoptosis: activation of caspases-8 and -9 and inactivation of PARP. The species retained after the model reduction are colored. **B.** The modular view of the model. The dashed connections were deleted during the model reduction. **C.** The minimal reduced model. **D.** The graphical notation used for representation of the models A-C.

### Reduction of the CD95-signaling model

We started the Bentele's model reduction by excluding the direct dependence of the virtual species $x_{aa}$ on caspases-6 and -7. For this purpose, we approximated the amount of caspases-7 by the linear function of caspase-3 concentration:

$$C_{casp7} \approx a \cdot C_{casp3}, \qquad \textbf{(12)}$$

where $a = 0.18$ for both activation scenarios. In addition, we considered inequalities holding for the parameters in the formula (10):

$$k_{7act} \cdot C_{casp7} \gg k_{36act} \cdot C_{casp3} > k_{36act} \cdot C_{casp6},$$

$$Km_{367act} \gg 1 - x_{aa},$$

and simplified it:

$$\frac{dx_{aa}}{dt} = \frac{(k_{36act} + 0.18 \cdot k_{7act})}{Km_{367act}} \cdot (1 - x_{aa}) \cdot C_{casp3} - k_{dd} \cdot x_{aa}. \qquad \textbf{(13)}$$

After that we decomposed the model into modules (fig. 3B) corresponding to three biological steps: activation of caspase-8, cytochrome *C*-induced activation of caspase-9 and PARP cleavage down-regulated by these caspases.

Below we provide detailed description of the reduction process of all modules. The process is based on the flow chart represented in the figure 2. If it is enough for a reader to have a general idea of model reduction without going into full details, we suggest to go directly to the last two paragraphs of this section, where you find the summary of the model reduction procedure.

*Caspase-8 activation module* includes 20 species and 14 reactions (besides the reactions of degradation), which could be divided into three groups: cleavage of procaspase-8 at the DISC complex, its activation triggered by caspase-3, and inhibition of complexes containing DISC by cFLIPL and cFLIPS (Table 1, fig. 3A, reactions br1-br14).

Firstly, we eliminated slow reactions br12 and br13 according to the method MR1. Next, we applied the quasi-steady-state analysis (MR2) to the module and removed quasi-stationary intermediates CD95R:CD95L, DISC:pro8, DISC:$pro8_2$, DISC:p43/p41 and DISC:cFLIPL. Thus, in particular, we got two main reactions instead of br1-br6: fast formation of the DISC complex (Table 1, fig. 3B, br1*) and slower activation of caspase-8 in this complex (br2*).

Further, we noticed that the consumptions of cFLIPL and cFLIPS satisfied the same kinetic laws. Therefore, these species could be lumped (MR3) resulting in the reaction br4*:

$$\text{cFLIP} + 2\cdot\text{DISC} + \text{pro8} \rightarrow \text{cFLIP:DISC:pro8},$$

where cFLIP indicated two isoforms cFLIPL and cFLIPS so that $C_{cFLIP} = C_{cFLIPL} = C_{cFLIPS}$.

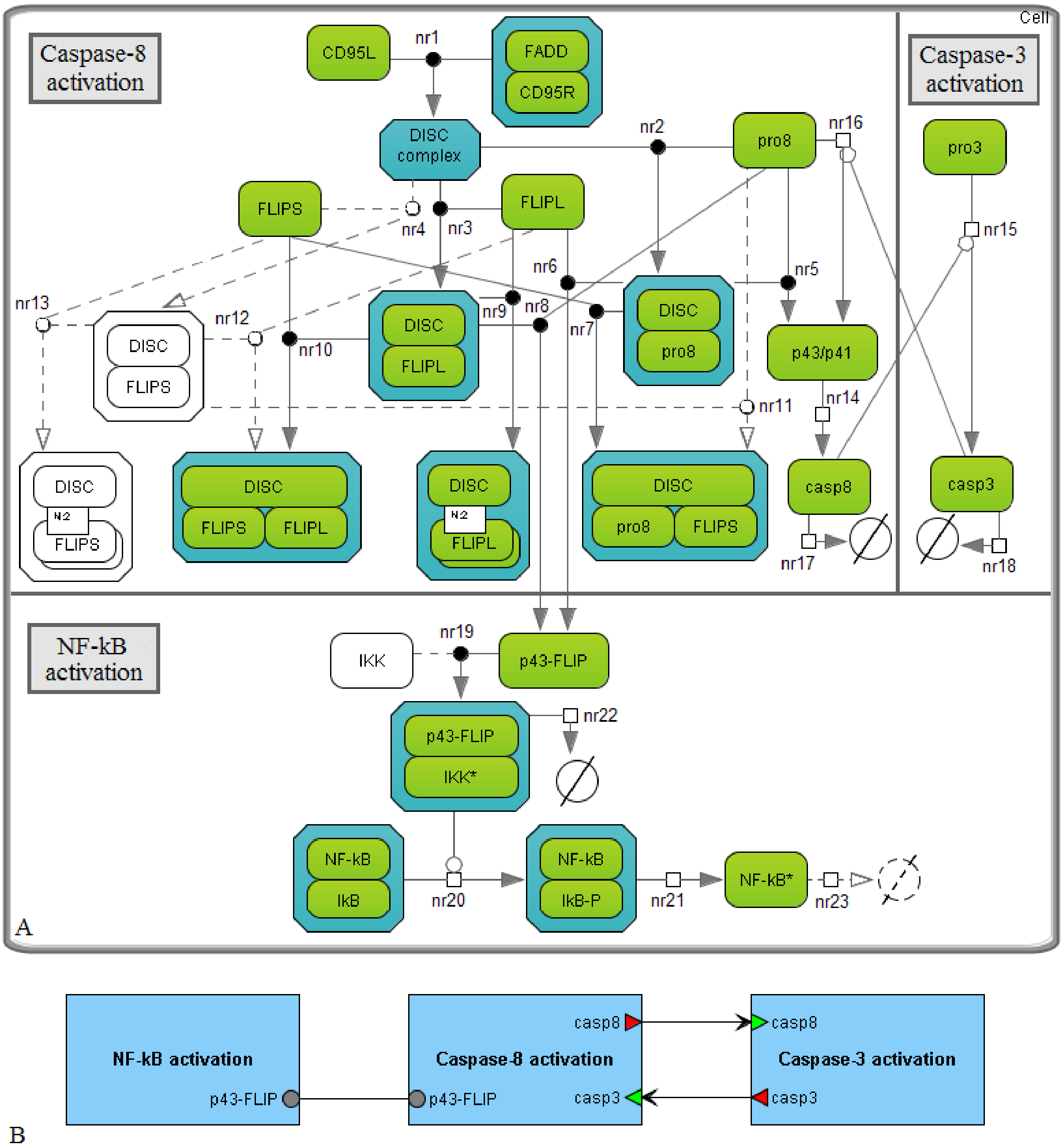

**Figure 4 – The original model by Neumann et al. and results of the model reduction**

**A.** The original model decomposed into modules according to three steps of the CD95 signaling pathways: activation of caspase-8, pro-apoptotic pathway resulting in caspase-3 activation and anti-apoptotic pathway regulated by NF-κB. The species retained during the model reduction are colored; reactions are represented by solid lines. **B.** The modular view of the model. Activation of caspase-8 and p43-FLIP (product of cFLIPL cleavage) occurs at the DISC complex and triggers simultaneous processes of cell death and survival.

We approximated the concentration of caspase-6 by the linear function $b \cdot C_{casp3}$, $b = 0.145$ (MR4), and merged reactions br7 and br8 of caspases-8 activation induced by caspases-3 and mediated by caspases-6 into one reaction br3*:

pro8 –casp3 → casp8.

Since the reaction br8 followed the Michaelis-Menten kinetics with the constants $k_{68}$ and $Km_{68}$, then the reaction rate of br3* was provided by the kinetic law

$$v_{br3^*} = k_{38} \cdot C_{pro8} \cdot C_{casp3} / \left(Km_{38} + C_{pro8}\right),$$

where $k_{38} = 0.145 \cdot k_{68}$ and $Km_{38} = Km_{68}$. Analysis of this kinetic law for low level of CD95L ensured that $C_{pro8} \gg Km_{38}$. In addition, in the case of the fast activation scenario the value of $v_{br3^*}$ was much lower than the rate of caspase-8 activation mediated by CD95. Thus, we established $v_{br3^*} = k_{38} \cdot C_{casp3}$ without significant changes in the results of the model simulation (MR1, MR5).

Complementing the above reduction steps, we removed the elements that were unnecessary to fit the experimental dynamics provided by Bentele *et al.* These elements were degradation reactions of CD95L, CD95R and FLIP, and the blocked DISC complex (inactivated DISC). As a result, we introduced the conservation law:

$$C_{CD95L} - C_{CD95R} = L_0 - R_0 = const,$$

where $L_0$ and $R_0$ denote initial concentrations of the ligand and receptor, respectively. Solving the differential equation

$$\frac{dC_{CD95R}}{dt} = -k_{LR} \cdot C_{CD95R} \cdot C_{CD95L},$$

we obtained the analytical function for the receptor concentration:

$$C_{CD95R} = \frac{R_0 - L_0}{1 - L_0/R_0 \cdot \exp(-k_{LR} \cdot (R_0 - L_0) \cdot t)} \quad \textbf{(14)}$$

*The module of caspase-9 activation* consisted of 19 species and 14 reactions (besides the reactions of degradation) including activation of caspase-9 triggered by cytochrome *C*, cleavage/activation of Bid by caspases-2 and -8, IAP inhibition of caspase-9 activity and binding of inhibitors by Smac (br15-br26, br35, br38). Based on the method MR1, we eliminated the last two reversible reactions, as well as the slow reaction br20 of the apoptosome complex dissociation. Further, since this complex was a quasi-stationary intermediate and the rate of br15 was linearly dependent on a difference of br16 and br17 rates, we reduced the chain of procaspase-9 activation (br15-br20), using the methods MR2 and MR4, to the form:

$$\begin{aligned}&\text{Cyt C stored} \rightarrow \text{Cyt C:Apaf-1},\\&\text{pro9 -Cyt C:Apaf-1} \rightarrow \text{casp9}.\end{aligned} \qquad \textbf{(15)}$$

Accordingly to Bentele *et al.*, we have the following rule of cytochrome $C$ release from mitochondria:

$$C_{Cyt\ C\ stored} = C_{Cyt\ C\ stored}(0) \cdot f_{release}(t),$$

$$\frac{dC_{Cyt\ C\ released}}{dt} = -\frac{dC_{Cyt\ C\ stored}}{dt}$$

Here $f_{release}(t)$ satisfies the formula (11). Taking into account (15) and ignoring degradation of the complex Cyt C:Apaf-1, we got

$$\frac{dC_{Cyt\ C:Apaf-1}}{dt} = -c \cdot \frac{dC_{Cyt\ C\ stored}}{dt}$$

where $c = 0.59$ was the coefficient of linearity.

Then, we noticed that the experimental measurements of procaspase-9 concentration were presented by Bentele and his colleagues only for 200 ng/ml of anti-CD95. In this case, degradation of procaspase-9 was insignificant. Thus, neglecting it and considering the kinetic law $k_{Apop} \cdot C_{pro9} \cdot C_{Cyt\ C:Apaf-1}$ of the second reaction in (15), we obtained the differential equation of procaspase-9 dynamics:

$$\frac{dC_{pro9}}{dt} = -0.59 \cdot \left(1 - f_{release}(t) \cdot C_{Cyt\ C\ stored}(0) \cdot k_{Apop} \cdot C_{pro9}\right).$$

Solving it, we found:

$$C_{pro9} = C_{pro9}(0) \cdot \left(\frac{f_{release}(t) \cdot (\exp(k_{contr} \cdot t) - 1) + 1}{\exp(k_{contr} \cdot t)}\right)^{\frac{0.59 \cdot C_{Cyt\ C\ stored}(0) \cdot k_{Apop}}{k_{contr}}}. \qquad \textbf{(16)}$$

Finally, analyzing reactions of Bid activation (br21-br23), we removed the slower reaction mediated by caspase-2. The similar reaction involving caspase-8 followed the Michaelis-Menten kinetics with the constant $Km_{28Bid} \gg C_{Bid}$. Therefore, we redefined the kinetics of this reaction based on the law of mass action (MR5):

$$v_{br7^*} = k_{8Bid}/Km_{8Bid} \cdot C_{casp8} \cdot C_{Bid}.$$

*The module of PARP inactivation* contained 13 species (including the virtual species $x_{aa}$) and 11 reactions (excepting reactions of degradation). Six reactions (activation of caspases-3 and -7 by caspases-8 and -9, and cleavage/inactivation of PARP) were based on the Michaelis-Menten kinetics (br27-br32), four reactions reproduced the reversible inhibition of caspases-3 and -7 based on the law of mass action (br33, br34, br36, br37), and one corresponded to the production of $x_{aa}$ (br38), simplification of which was discussed above.

We deleted slow reactions of casp3:IAP and casp7:IAP complexes dissociation and ignored slow degradation of procaspase-7 and IAP. Then analyzing the cleavage of

procaspases-3 and -7, we eliminated reactions triggered by caspase-9, which were slower in comparison to the similar reactions induced by caspase-8. For the same reason we ignored the reaction of PARP inactivation by caspase-3 (MR1).

Considering the remaining reactions with the Michaelis-Menten kinetics (br29, br32), we used the following inequalities $Km_{78} \gg C_{pro7}$ and $Km_{376act} \gg C_{PARP}$, which, in combination with MR5 and (12), led to the following kinetic laws:

$$v_{br29} = \frac{k_{78}}{Km_{78}} \cdot C_{casp8} \cdot C_{pro7}, \qquad v_{br32} = \frac{k_{3act}}{Km_{367act}} \cdot C_{casp3} \cdot C_{PARP},$$

where $k_{3act} = 0.18 \cdot k_{7act}$.

Thus, the CD95-signaling model consisting of 43 species (including one virtual species), 80 reactions and 45 kinetic parameters was approximated by a model with 18 species, 26 reactions and 25 parameters, except the constant concentration of Cytochrome *C* (fig. 3C). Figure 5 shows the comparison of simulated concentrations of these models with experimental dynamics obtained by Bentele *et al*. for the 11 species mentioned above. Since this dynamics was expressed in arbitrary units, we uniquely translated it into precise values for all proteins with non-zero initial values. However, for such molecules as caspase-8, p43/p41, tBid and cPARP, precise values directly depended on the concentration levels at the end or in the middle of time series. Accordingly, we computed these levels during the model reduction and recalculated experimental dots if necessary.

Species without experimental evidence (CD95L, DISC, cFLIP, DISC:pro8:cFLIP, IAP, caspase-3 and virtual species $x_{aa}$) and reactions br1*-br13* directly affect the simulated dynamics of the experimentally measured concentrations (Table 2). Thus, further reduction of the model by removing these elements is impossible. Regarding the degradation process, we agree with Bentele *et al.* that experimental data cannot be matched if this process is fully ignored. Actually, if we remove any of the retained reactions of degradation, then excepted concentration dynamics will be impaired. Thus, we found the minimal complexity approximation of the original model.

### Reduction of the CD95-mediated and NF-κB signaling model

As was mentioned above, Neumann *et al.* simplified their model in order to reduce the large number of free parameters. The authors noted that further simplification of the model was not possible because the model fit significantly decreased. However, we removed slow reactions (nr4, nr11-nr13, fig. 4A) using the method MR1, and removed the reaction of NF-κB degradation (nr23), which was not significant for reproducing the experimental data of the original model. Note that we retained the slow reactions nr6 and nr7, the former of which regulates apoptosis inhibition in the case of high levels of cFLIPL and procaspase-8 in accordance with the precursor model prediction (see the analysis of the predictions below), and the latter of which will be required to combine this model with the Bentele's model hereafter.

Based on the method MR6, we also took into account the inequality $C_{IKK} \gg C_{p43-FLIP}$ and simplified the kinetic law

$$v_{nr19} = k_{p43-FLIP_IKK} \cdot C_{IKK} \cdot C_{p43-FLIP}$$

of the reaction nr19 to the form

$$v_{nr19^*} = k_{p43-FLIP_IKK} \cdot C_{IKK}(0) \cdot C_{p43-FLIP}.$$

Therefore, we reduced the number of model species from 23 to 20, the number of reactions from 23 to 18 and the number of kinetic parameters from 17 to 15 (besides the constant concentration of IKK). These modifications did not change the fit to the experimental data provided by Neumann *et al.* (Figure 6).

### Model composition

Analysis of the reduced models based on the Akaike criterion (7) confirmed that they had lower complexity than the original models (table 3). The difference between the mean $AIC$ coefficients (8) of the models decreased by 60%, relative to the initial value. In addition, the reduced Neumann's model better approximated experimental data. Therefore, we used it as the basis for the model composition.

We took into account that the considered experimental data were obtained with SKW 6.4 [18] and HeLa [19] cells, which were shown to behave as type I [21] and type II [28] cells, respectively. In this regard, we assumed that:

(A) initial species concentrations could vary for different cell lines;

(B) kinetic parameters of all reactions (besides the degradation rate modeled as the function $k_d \cdot x_{aa}^2$) have the same values for both cell lines, whereas the value of $k_d$ could be regulated by various entities and, moreover, is dependent on the initial concentration of anti-CD95 [18];

(C) type I and type II cells conform to different reaction chains of caspases-3 activation [21].

We constructed the composite model of CD95 and NF-κB signaling in the following way. Firstly, according to the assumption (C), we replaced reaction nr15 in the Neumann's model by a chain of three reactions:

$$\begin{aligned} &\text{Bid –casp8} \rightarrow \text{tBid}, \\ &\text{pro9 –tBid} \rightarrow \text{casp9}, \\ &\text{pro3 –casp9} \rightarrow \text{casp3}. \end{aligned} \qquad \textbf{(17)}$$

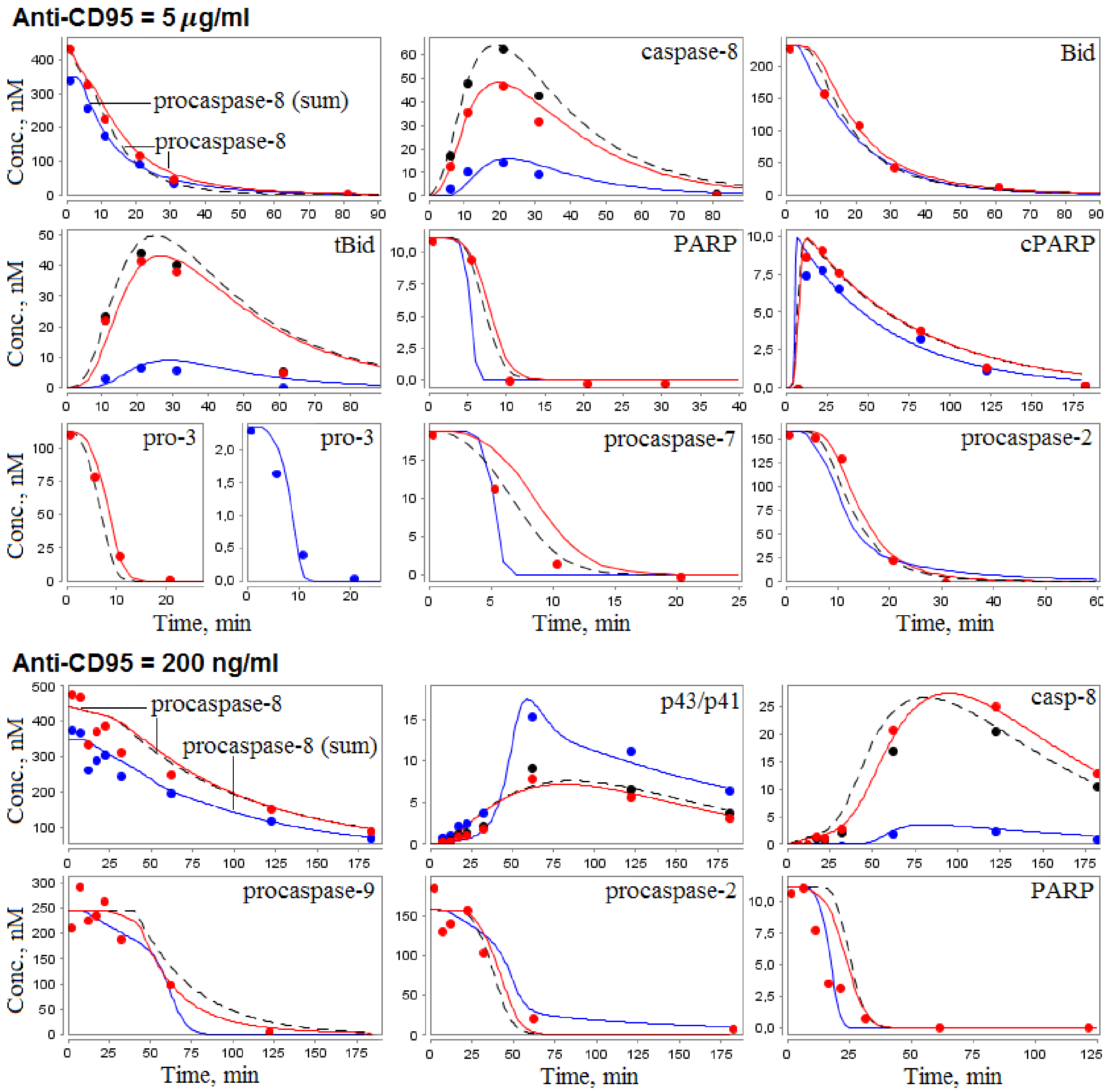

**Figure 5 – Results of the composite model approximation to the experimental data by Bentele *et al*.**

The experimental data (dots) were obtained by Bentele *et al.* [18] using Western blot analysis and expressed as relative values. Once simulated values (curves in the figure) were found for the original Bentele's model (solid red), the reduced model (dashed black) and the composite model (solid blue), experimental points were recalculated taking into account these values. Wherever blue or black dots are missing, they coincide with the red dots. Note that considering the concentration of procaspase-8 in the original model, we observed experimental dynamics for the separate species only, but not for the total amount as expected from experiments. Thus, validating parameters of the composite model, we evaluated the total amount of procaspase-8.

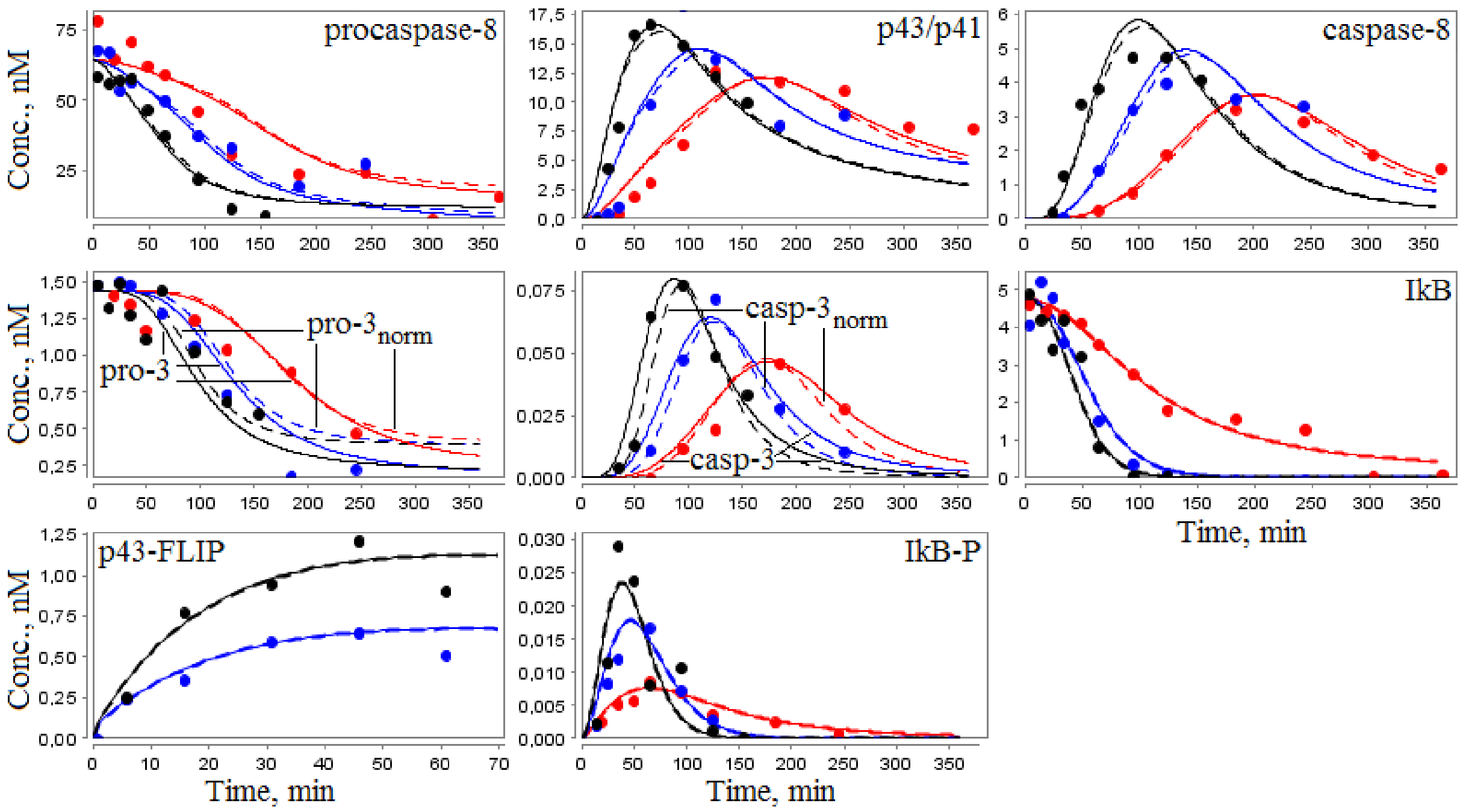

**Figure 6 – Comparison of the composite model simulation results with the experimental data by Neumann *et al*.**

The simulated concentrations of the original and reduced models (solid lines), as well as the simulated values of the composite model (dashed lines) were obtained by Neumann *et al.* [19] for three different concentrations of anti-CD95: 1500 ng/ml (black), 500 ng/ml (blue), and 250 ng/ml (red). We did not recalculate the relative experimental data (dots) as in the case of the Bentele's model due to a slight difference between the results of the composite and precursor models. An exception was made for values of procaspase-3 and caspase-3, which were normalized for the composite model by the term 0.1 for convenience of the visual plots comparison.

The first reaction of this chain had the same kinetic law as br7*. The other reactions represented br6* and nr15 with kinetics modified using the law of mass action. Such modification of br6* was necessary to get the slow increase of caspase-3 concentration experimentally described by Neumann *et al.* [19]. Modification of nr15 consisted in substitution of $C_{casp9}$ for $C_{casp8}$ .

Secondly, we supplemented the Neumann's model with reaction of caspase-3 inhibition by IAP and evaluated parameters in this reaction together with parameters in (17) using the corresponding data by Neumann *et al.* and optimization tools of BioUML [29].

Thirdly, we analyzed the changes of the initial concentrations and parameters of the derived model required to reproduce the Bentele's experimental data fixing levels of

procaspases-3, -8, cleaved product p43/p41, and caspase-8 (table 4). All the changes, except step 4, agreed with assumptions (A) and (B). To reproduce the species dynamics, we had to increase the initial concentration of procaspase-3 in the case of HeLa cells by an order of magnitude.

Finally, we combined the modules of caspase-8 and NF-κB activation with the modules of caspase-9 activation and PARP inactivation (fig. 7). For this purpose, we modified the last two modules based on (17) and refitted their parameters as well as the concentrations mentioned in table 4, using experimental data by Bentele *et al.* For better fit we also supplemented the model by degradation reactions of some species (procaspase-9, caspase-9, DISC:pro8 and DISC:pro8:FLIPS).

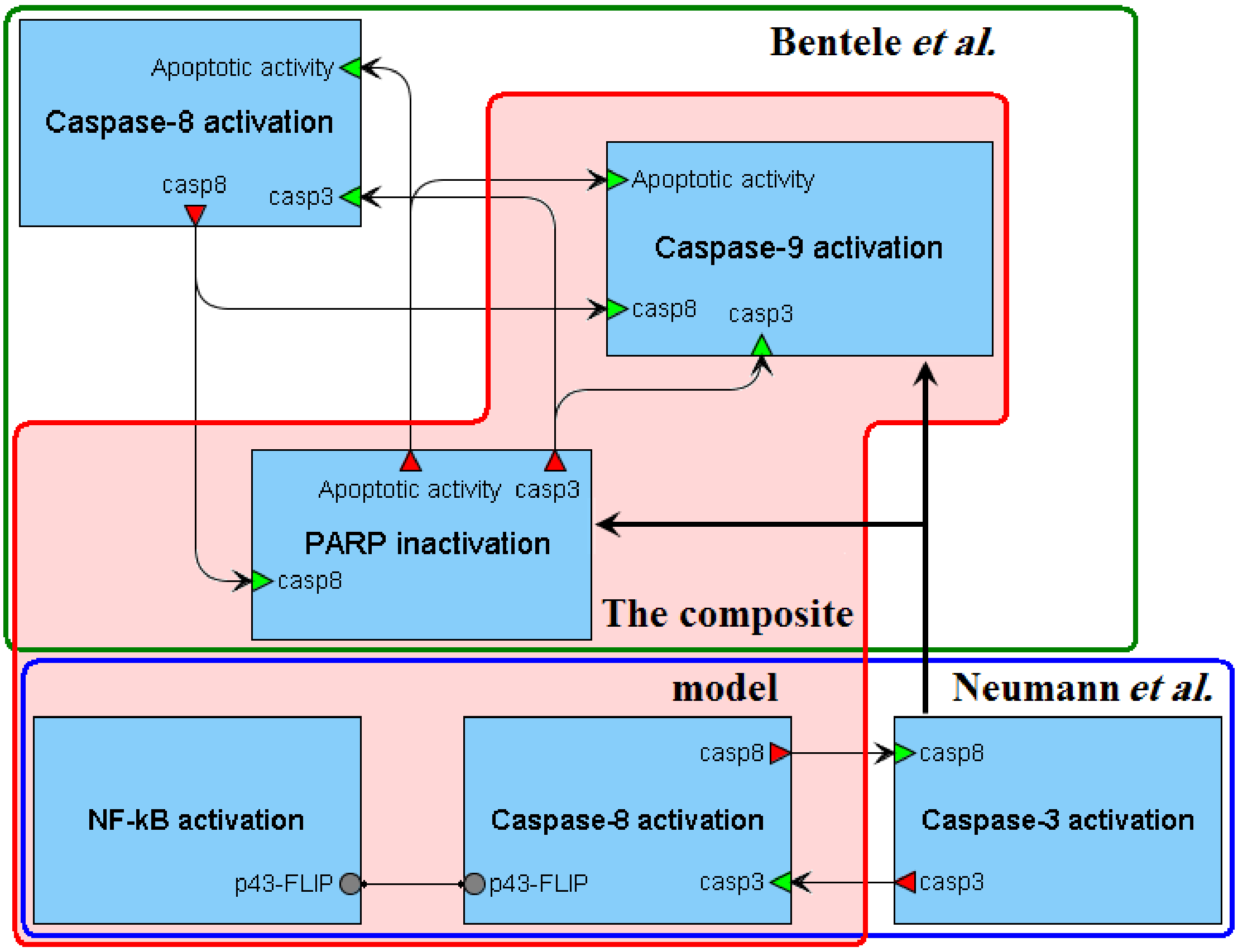

**Figure 7 – Creation of the composite model**

The modules of caspase-8 and NF-κB activation were taken from the reduced Neumann's model and combined with the modules of caspase-9 activation and PARP inactivation isolated from the reduced Bentele's model. In addition, two last modules were supplemented by reactions from the modified module of caspases-3 activation.

The resulting composite model (fig. 8) consists of 30 species (including one virtual species), 38 reactions (including 15 reactions of the species degradation) and 30 kinetic parameters, except for the constant concentration of IKK. The model showed a good fit between the simulation results and the experimental data (fig. 5, 6).

Calculation of the mean $AIC$ coefficient for this model revealed that it has the same level of complexity as the reduced models (Table 3). We also noted that reduction of both models did not significantly alter the steady-state concentrations (Tables 5S-6S). However, the composite model has different steady-states, which are more stable according to the mean sensitivities (9) in the cases of SKW 6.4 cells and HeLa cells stimulated with 250 ng/ml of anti-CD95.

**The reduced models are equivalent to the precursor models with respect to their biological predictions**

Analyzing the models constructed by Neumann *et al.* and Bentele *et al.*, we divided predictions made by them into two groups: qualitative and quantitative. The first group concerned the model network and described mechanisms of the protein-protein interactions. It is this group to which we assigned the experimentally validated Neumann's predictions that processed caspases are not required for NF-κB activation, as well as that pro-apoptotic and NF-κB pathways diverge already at DISC. The corresponding network of reactions was preserved in the reduced model. Thus, the prediction remained valid.

The second group of predictions characterized behavior of the models after changing concentrations of some species, such as CD95L, CD95R, procaspase-8, and inhibitors cFLIPL, cFLIPS, and IAP. Analyzing the predictions of the models by Neumann *et al.* and Bentele *et al.* listed in the tables 5 and 6 respectively, we concluded that the reduction of the first of them did not alter the species dynamics. For the second reduced model, we detected some minor changes in the predictions. However, in general, the model behavior remained in a good agreement with the Bentele's experiments.

**Model composition modifies some properties of the precursor models and leads to new predictions**

Considering the predictive ability of the composite model, we found that in the case of HeLa cells it was completely preserved (table 5).

In the case of SKW 6.4 cell line, we estimated the value of degradation rate parameter $k_d$ based on the experimental observations by Bentele *et al.* for 1-10 ng/ml of anti-CD95 (table 6). When performing simulation experiments of the model for this cell line, we used three different values of $k_d$ depending on whether the initial concentration of CD95L was within the range 1-100 ng/ml, within the range 100-1000 ng/ml or greater than 1000 ng/ml (table 3S).

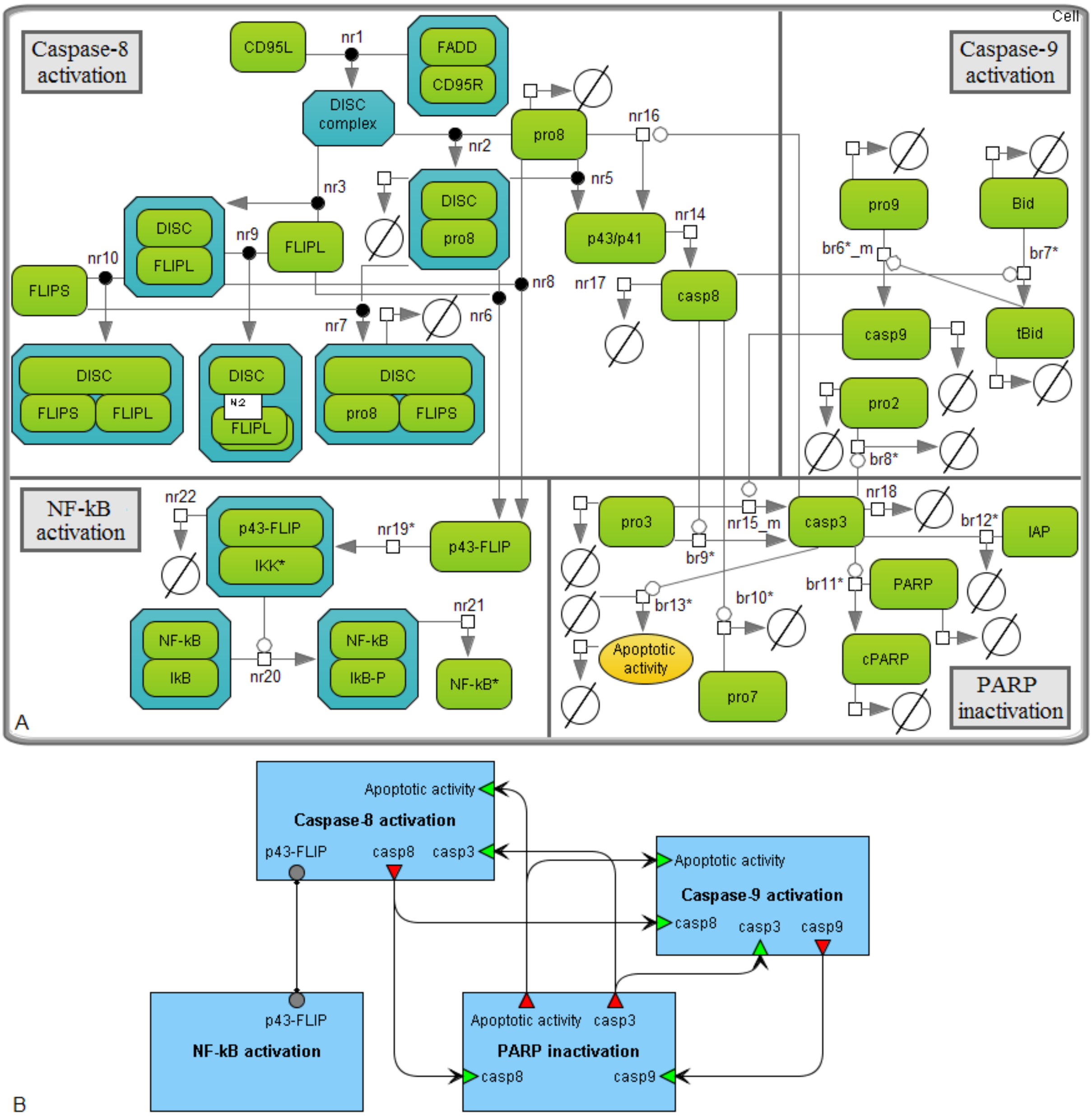

**Figure 8 – The composite model of the CD95- and NF-κB-signaling**

**A.** The model integrating pro- and anti-apoptotic machinery described by the models by Neumann *et al.* and Bentele *et al.* Reactions with the "nr" titles were taken from the first model, while the "br" titles indicate reactions from the second one. The "*" symbol marks reactions defined in the reduced models. The subscript "m" means that reaction was modified for the models composition. **B.** The modular view of the model. Activation of caspase-8 triggered by CD95 leads to NF-κB activation and cell survival, on the one hand, and PARP cleavage resulted in the cell death, on the other hand.

Analyzing the model predictions, we detected differences in the behavior of the Bentele's and the composite models (table 6). The first of them describes a threshold mechanism for CD95-induced apoptosis implying that this process is completely stopped when CD95L concentration is below the critical value. However, the model predicts a sharp increase of cPARP immediately when the level of CD95L exceeds the threshold (table 6, № 1). In contrast, the composite model shows a gradual increase of cPARP with increase of the ligand level. To determine the apoptotic threshold in this case, we found concentration of CD95L for which cPARP ratio is equal to 10% of the initial PARP amount [30]. As in the original model, this concentration was in the range of 1-10 ng/ml. Additionally, the models revealed sensitivity of the threshold to the concentration of cFLIP (table 6, № 2). Both of them predicted the cell death phenotype upon stimulation by 1 ng/ml anti-CD95 and the level of cFLIP decreased by ~50%.

The next prediction, which we observed for SKW 6.4 cells, considers a delay of caspase-8 activation caused by low concentrations of anti-CD95 (table 6, № 3). The maximal delay predicted by the Bentele's model was about 2500 min, whereas the composite model detected a much lower value (700-800 min). Analysis of dependence of caspase-3 concentration on initial values of IAP and CD95L (table 6, № 4) also demonstrated differences in the models' predictions. Thus, low level of IAP (less than 1 nM) in the original model results in complete cell death. However, under the same condition the composite model shows a smooth increase of caspase-3 levels with CD95L changing from 0.3 nM to 1000 nM. If the amount of the ligand is less than 0.3 nM, IAP blocks apoptosis completely. In addition, the Bentele's model predicts that high concentration of IAP prevents a significant increase of caspase-3 even for high levels of the ligand, whereas in the composite model high concentration of CD95L leads to cell death.

**Limitations of the composite model**

We analyzed a set of parametric constraints which allowed us to reduce the precursor models (table 10S). Following them, we can recover a detailed description of the investigated process. Connection of the composite model with the original model by Neumann *et al.* was preserved, whereas with the original model by Bentele *et al.* it was partially lost since the chains of species interactions were changed for the sake of combining the models together. In particular, the pathway of caspase-8 activation in DISC was derived from the Neumann's model as well as the reaction chain of procaspase-9 cleavage was modified. However, following the constraints of table 10S, we can complete the composite model with such reactions as Bid truncation by caspase-2 and PARP cleavage triggered by caspase-7. We can also recover the details of caspase-9 activation mediated by cytochrome *C* and Smac release from mitochondria but with altered kinetics of release. For now, the question of admissibility of such modifications as well as extension of the composite model based on other apoptosis models remains open and is a challenge for further research.

# Discussion

In this paper, we considered two approaches to development of mathematical models of cell fate decisions. The first concerns the methodology of model reduction and involves approximation of one model by another one of lower dimension without affecting dynamics of experimentally measured species. The second implies composition of the models and aims at reproducing experimental dynamics of all precursor models.

There are many advantages of working with low-dimensional models [8]. In particular, the researcher has a clear vision of the most important biochemical reactions taking place in the modeled system as well as better understanding of them. Low-dimensional models are easier to analyze and faster to simulate. This helps to save time and enhances productivity. The main limitation of model reduction consists in loss of biological information. However, it should be noted that even if some information (for example, about very slow biochemical reactions) may be lost, it can result in a more clear understanding of the most important interactions and allows focusing on the decisive processes in the model, predictive ability of which is reasonably preserved. Hence, this limitation can be transformed into an advantage.

Model composition aiming at getting a single model from several ones is useful because in such a case a computational biologist is able to investigate the composite model behavior under different conditions that cannot be performed in the precursor models separately. For example, our model allows studying the role of p43-FLIP or IAP in the type I SKW 6.4 or type II HeLa cells, respectively, that might become a task for the future work. In other words, we constructed the model that describes pro- and anti-apoptotic signal transduction in different cell types with reasonable accuracy instead of a couple of different models. Whenever necessary, some reaction chains and parameters can be switched off giving opportunity to simulate a certain type of cells. In addition, the composite model covers experimental data obtained from all precursor models, each of which separately satisfies its own data only.

Model composition sometimes causes modification of some properties of the initial models, resulting in new testable predictions. In our case, such predictions were related to SKW 6.4 cells, and some simulation results were different from the corresponding results of the original Bentele's model. Nonetheless, the composite model behavior remained in good agreement with experimental data used in modeling by Bentele *et al*. Thus, the question of what model (original or composite) is more realistic requires further experimental investigation.

One of the questions that can be asked concerns the need of reducing models in order to combine them. An attempt to construct the composite modular model of apoptosis was made in [26]. This attempt revealed multiple conceptual and methodological difficulties. The main obvious obstacles included:

- choice of elements in the overlapping parts of different models;
- incomparable sets of parameters;
- the lack of experimental data, as well as inability to use the data obtained for different cell lines or in different ways (e.g. single-cell or cell culture measurements);
- inability to make accurate predictions.

Model reduction allows solving some of these problems. In particular, reducing the number of model elements, we reduce the overlapping parts as well. This may be essential for direct combining of models. In addition, the complexity of the reduced model become comparable with the complexity of available experimental data. Therefore, the risk of model overfitting is decreased.

In the case when two models are directly related (for example, one model was emerged from the other), their composition may be significantly easier in comparison to composition of quite different models. In our work, the models by Bentele *et al.* and Neumann *et al.* are not directly related, as it could be expected as these models were constructed in the same research group. For example, they use different reaction chains of caspase-8 activation and different values of kinetic parameters in the overlapping reactions.

Another quite useful principle that we used in modeling was modular structure of the developed model. This principle provides flexibility for future extensions. Thus, we are planning to extend the composite model and supply it with modules and data from studies of TRAIL [31-33] and TNF-α [34, 35] signaling, apoptosome-dependent caspase activation [36] and p53 oscillation system [37]. It is noteworthy that the model by Laussmann *et al.* [33], describing TRAIL-induced activation of caspase-8, emerged from the original study by Bentele *et al.* [18]. This fact may help us in our future work.

Characterising the models used in our work, we can say that we saved our time working with the model by Neumann *et al.* [19] derived from the BioModels database [27], and spent a lot of time to reproduce the model by Bentele *et al.* [18] that had not existed in a ready-to-use format. Thus, we would like to emphasize that for the common benefit, the systematic application of biological standards in modeling (e.g. SBML [38] and SBGN [39]) and depositing the working models in public databases (e.g. BioModels [27]) would significantly facilitate analysis of existing biomathematical models and using them as the base for development of new composite systems.

Finally, we believe that the composite model itself may be useful for further investigation of apoptosis.

# Conclusions

Mathematical modeling provides a powerful tool for studying the properties of biological processes. Methods of model reduction allowed us to take a first step towards validation of the modular model of apoptosis [26]. Using these methods, we composed two models describing pathways of CD95- and NF-κB-signaling in one without affecting the fit to the experimentally measured dynamics and model predictions. For the model reduction and composition, we used the BioUML software that was extended by the required methods of analysis.

The models by Bentele *et al.* and Neumann *et al.* reconstructed in BioUML and represented using SBML format and SBGN notation, as well as their reduced and decomposed versions are available at http://ie.biouml.org/bioumlweb/#de=databases/The%20composite%20model%20of%20CD95%20and%20NF-kB%20signaling.

# Competing interests

The authors declare that they have no competing interests.

# Authors' contributions

EK performed reduction, integration and analysis of the apoptosis models. AZ coordinated the model reduction part of the study. RS and FA coordinated the study of the apoptosis machinery. FA also led implementation of the analysis toolbox in the BioUML software. All authors have participated in writing and editing the manuscript, as well as read and approved the final manuscript.

## Acknowledgements

This work was supported by FP6 grant 037590 "Net2Drug", FP7 grant 090107 "LipidomicNet", EU FP7 (project APO-SYS) and Agence Nationale de la Recherche (project ANR-08-SYSC-003 CALAMAR). AZ is a member of the team “Systems Biology of Cancer,” Equipe labellisée par la Ligue Nationale Contre le Cancer.

# Tables

## Table 1 – Summary of reactions from the original model by Bentele *et al*. and the reduced model

| Original model | | | Reduced model | |
|---|---|---|---|---|
| **№** | **Reactions (Kinetics)** | **Rates (nM/min)** | **№** | **Reactions (Kinetics)** |
| ***Activation of caspase-8 induced by CD95*** | | | | |
| br1 | CD95L + CD95R → CD95R:CD95L (MA) | $10^{1}/10^{0}$ | br1* | CD95L → DISC ($k_{LR} \cdot C_{CD95R} \cdot C_{CD95L}$, $C_{CD95R}$ is determined by the formula (14)) |
| br2 | FADD + CD95R:CD95L → DISC (MA) | $10^{1}/10^{0}$ | | |
| br3 | pro8 + DISC → DISC:pro8 (MA) | $10^{0}/10^{-1}$ | | |
| br4 | pro8 + DISC:pro8 → DISC:$pro8_2$ (MA) | $10^{0}/10^{-1}$ | | |
| br5 | DISC:$pro8_2$ → DISC:p43/p41 (MA) | $10^{0}/10^{-1}$ | br2* | pro8 –DISC → casp8 ($k_{DISC_pro8} \cdot C_{pro8} \cdot C_{DISC}$) |
| br6 | DISC:p43/p41 → casp8 + DISC (MA) | $10^{0}/10^{-1}$ | | |
| ***Activation of caspase-8 by caspase-3*** | | | | |
| br7 | pro6 -casp3 → casp6 (M-M) | $10^{0}/10^{-1}$ | br3* | pro8 –casp3 → casp8 ($k_{38} \cdot C_{casp3}$) |
| br8 | pro8 -casp6 → casp8 (M-M) | $10^{-1}/10^{0}$ | | |
| ***Inhibition of the DISC complex*** | | | | |
| br9 | DISC + cFLIPL → DISC:cFLIPL (MA) | $10^{1}/10^{0}$ | br4* | cFLIP + 2·DISC + pro8 → DISC:FLIP:pro8 ($k_{DISC_FLIP} \cdot C_{FLIP} \cdot C_{DISC}$) |
| br10 | pro8 + DISC:cFLIPL → DISC:cFLIPL:pro8 (MA) | $10^{0}$ | | |
| br11 | DISC + cFLIPS → blocked DISC (MA) | $10^{1}/10^{0}$ | | |
| br12 | DISC:pro8 + cFLIPL → DISC:cFLIPL:pro8 (MA) | $10^{-2}/10^{-3}$ | br5* | DISC:FLIP:pro8 → p43/p41 ($k_{DFp8} \cdot C_{DISC:FLIP:pro8}$) |
| br13 | DISC:pro8 + cFLIPS → blocked DISC (MA) | $10^{-2}/10^{-3}$ | | |
| br14 | DISC:cFLIPL:pro8 → p43/p41 + blocked DISC (MA) | $10^{-1}$ | | |
| ***Activation of caspase-9 triggered by cytochrome C*** | | | | |
| br15 | Cyt C stored → Cyt C ($f_{release}(t) \cdot C_{Cyt\ C\ stored}(0)$) | $10^{2}$ | br6* | pro9 → cleavage (the formula (16)) |
| br16 | Apaf-1 + Cyt C → Cyt C:Apaf-1 (MA) | $10^{2}$ | | |
| br17 | Cyt C:Apaf-1 → Apaf-1 + Cyt C (MA) | $10^{1}$ | | |
| br18 | pro9 + Cyt C:Apaf → Apop (MA) | $10^{0}$ | | |
| br19 | Apop -casp3 → casp9 + Cyt C:Apaf-1 (M-M) | $10^{0}$ | | |
| br20 | Apop → casp9 + Cyt C:Apaf-1 (MA) | $10^{-2}$ | | |
| ***Activation of Bid*** | | | | |
| br21 | pro2 –casp3 → casp2 (M-M) | $10^{0}$ | br7* | Bid –casp8 → tBid ($\frac{k_{8Bid}}{Km_{8Bid}} \cdot C_{casp8} \cdot C_{Bid}$) |
| br22 | Bid –casp8 → tBid (M-M) | $10^{0}$ | | |
| br23 | Bid –casp2 → tBid (M-M) | $10^{-1}/10^{0}$ | br8* | pro2 –casp3 → cleavage ($\frac{k_{32} \cdot C_{casp3} \cdot C_{pro2}}{C_{pro2}+Km_{32}}$) |
| ***Blocking of IAP by Smac*** | | | | |
| br24 | Smac stored → Smac ($f_{release}(t) \cdot C_{Smac\ stored}(0)$) | $10^{2}$ | – | |
| br25 | Smac + IAP → IAP:Smac (MA) | $10^{-5}$ | | |
| br26 | IAP:Smac → Smac + IAP (MA) | $10^{-5}$ | | |
| ***Activation of caspases-3 and -7*** | | | | |
| br27 | pro3 –casp8 → casp3 (M-M) | $10^{1}/10^{0}$ | br9* | pro3 –casp8 → casp3 (M-M) |
| br28 | pro3 –casp9 → casp3 (M-M) | $10^{-3}/10^{0}$ | | |
| br29 | pro7 –casp8 → casp7 (M-M) | $10^{0}/10^{-1}$ | br10* | pro7 –casp8 → cleavage ($\frac{k_{78}}{Km_{78}} \cdot C_{casp8} \cdot C_{pro7}$) |
| br30 | pro7 –casp9 → casp7 (M-M) | $10^{-5}/10^{-3}$ | | |
| ***PARP inactivation*** | | | | |
| br31 | PARP -casp3 → cPARP (M-M) | $10^{-1}$ | br11* | PARP -casp3 → cPARP |
| br32 | PARP –casp7 → cPARP (M-M) | $10^{0}/10^{-1}$ | | |
| | | | | $(\frac{k_{3act}}{Km_{367act}} \cdot C_{casp3} \cdot C_{PARP})$ |
| ***Inhibition of caspases-3, -7 and -9*** | | | | |
| br33 | casp3 + IAP → casp3:IAP (MA) | $10^{0}$ | br12* | casp3 + IAP → inhibition (MA) |
| br34 | casp7 + IAP → casp7:IAP (MA) | $10^{-1}$ | | |
| br35 | casp9 + IAP → casp9:IAP (MA) | $10^{-3}$ | | |
| br36 | casp3:IAP → casp3 + IAP (MA) | $10^{-2}$ | | |
| br37 | casp7:IAP → casp7 + IAP (MA) | $10^{-4}/10^{-3}$ | | |
| br38 | casp9:IAP → casp9 + IAP (MA) | $10^{-5}/10^{-4}$ | | |
| ***Production of the virtual variable $x_{aa}$*** | | | | |
| br39 | $\rightarrow x_{aa}\ (1 - x_{aa}/Km_{367act} + (1 - x_{aa}) \cdot (k_{36act} \cdot C_{casp3} + k_{36act} \cdot C_{casp6} + k_{7act} \cdot C_{casp7}))$ | $10^{0}$ | br13* | $\rightarrow\ x_{aa}\ (\frac{k_{36act}+0.18 \cdot k_{7act}}{Km_{367act}} \cdot (1 - x_{aa}) \cdot C_{casp3})$ |

Prototypes of parameter notations in the original Bentele's model are listed in table 1S. Abbreviations used in the table represent MA – mass action kinetics and M-M – Michaelis-Menten kinetics. The column with rates includes rate orders for the fast/reduced activation scenarios.

**Table 2 – The role of species and reactions in the apoptotic process described by the reduced model**

| Species | Reactions | The role in the apoptosis process |
|---|---|---|
| CD95L | br1* | *Triggering the apoptosis process.* |
| DISC | br2* | *Activation of caspase-8 in the DISC complex.* In the case of a high ligand concentration, DISC is a slow species, so reactions br1* and br2* cannot be combined. |
| caspase-3 | br3* | *Activation of caspase-8 via the negative-feedback loop*. When the concentration of CD95L is reduced, the rate of br3* is an order of magnitude higher than the rate of br2*. Thus, activation of caspase-8 is mainly caused by caspase-3. |
| | br8* | *Cleavage/activation of procaspase-2* is confirmed by the experimental data. |
| | br9* | *Caspase-3 activation* plays a crucial role in the cleavage of procaspases-2, -8 and PARP. |
| | br11* | *PARP inactivation* is confirmed by the experimental measurements of PARP and cPARP concentrations. |
| cFLIP | br4* | *Inhibition of the DISC complex*. cFLIP blocks the activity of DISC preventing a significant increase of caspase-8 concentration. This conclusion is consistent with the observations by Bentele *et al.* asserting that down-regulation of cFLIP resulted in cell death occurring already upon low concentration of anti-CD95 (1 ng/ml). |
| DISC:cFLIP:pro8 | br5* | *Production of blocked p43/p41*, whose concentration is detected by the experimental data. |
| | br6*, br7*, br10* | *Activation of Bid and cleavage of procaspases-7 and -9.* These reactions determine dynamics of the mentioned species in agreement with the experimental data. |
| IAP | br12* | *Inhibition of caspase-3*. In the case of the reduced activation scenario, IAP prevents caspase-3 from reaching a significant level and, therefore, blocks significant increase of caspase-8 due to br3*. |
| $x_{aa}$ | br13* | The virtual variable $x_{aa}$ specifies the process of degradation. |

"Experimental data" in this table refers to the data obtained by Bentele *et al.* [18]

**Table 3 – Comparison of the investigated models according to Akaike's information criterion (*AIC*)**

| | Bentele *et al.* | | Neumann *et al.* | | Composite model |
|---|---|---|---|---|---|
| | Original model | Reduced model | Original model | Reduced model | |
| ***AIC*** | 182.46 | 138.83 | 201.58 | 198.17 | 318.10 |
| **Mean *AIC*** | 1.50 | 1.14 | 0.96 | 0.94 | 0.96 |

**Table 4 – Modifications of the reduced Neumann's model to reproduce the experimental data by Bentele *et al*.**

| Step | Changed parameters and concentrations | Initial values, modifications | Reason | Ranges and initial values of fitting |
|---|---|---|---|---|
| 1 | CD95R:FADD | 91.266 nM, increase | *Case: anti-CD95 = 5 µg/ml* Procaspase-8 is cleaved in about 30 minutes. This is possible only if DISC concentration reaches a level sufficient to accelerate the reaction nr2. | $[10^2, 10^3]$, 442.821, Bentele *et al.* |
| 2 | procaspase-8 | 64.477 nM, increase | *Case: anti-CD95 = 5 µg/ml* The concentration of caspase-8 should reach its maxim value in about 20 minutes. | $[10^2, 10^3]$, 442.821, Bentele *et al.* |
| 3 | FLIPS | 5.084 nM, increase | *Case: anti-CD95 = 200 ng/ml* As mentioned in table 2, activation of caspase-8 is mainly caused by caspase-3 and, therefore, is delayed approximately for 30 minutes. Since reactions nr2 and nr5 have the same rates, we cannot observe the delay, but when we reduce the rate of nr5, we can. The latter is achieved by upregulation of the DISC:pro8 complex inhibition. | $[10^1, 10^2]$, 65.021, Bentele *et al.* |
| 4 | k10 | 0.121 $nM^{-1}min^{-1}$, decrease by an order of magnitude | *Case: anti-CD95 = 200 ng/ml* Preventing a rapid growth of caspase-8 concentration. | 0.012 |
| 5 | procaspase-3 | 1.443 nM, refitting | *Case: anti-CD95 = 5 µg/ml* Improving approximation of the experimental data by Bentele *et al*. | $[10^0, 10^1]$, 1.443, Neumann *et al.* |
| 6 | Bid, procaspase-9 | 5.003 nM, 2.909 nM, replacement by Bentele's values | *Case: anti-CD95 = 200 ng/ml* Reproducing the experimental dynamics of procaspase-9. | 231.760 (Bid), 245.101 (pro-9), Bentele *et al.* |

**Table 5 – Analysis of predictions regarding apoptosis in HeLa cells as formulated by Neumann *et al*.**

| № | The composite model behavior | Predictions by Neumann *et al.* |
|---|---|---|
| 1* | 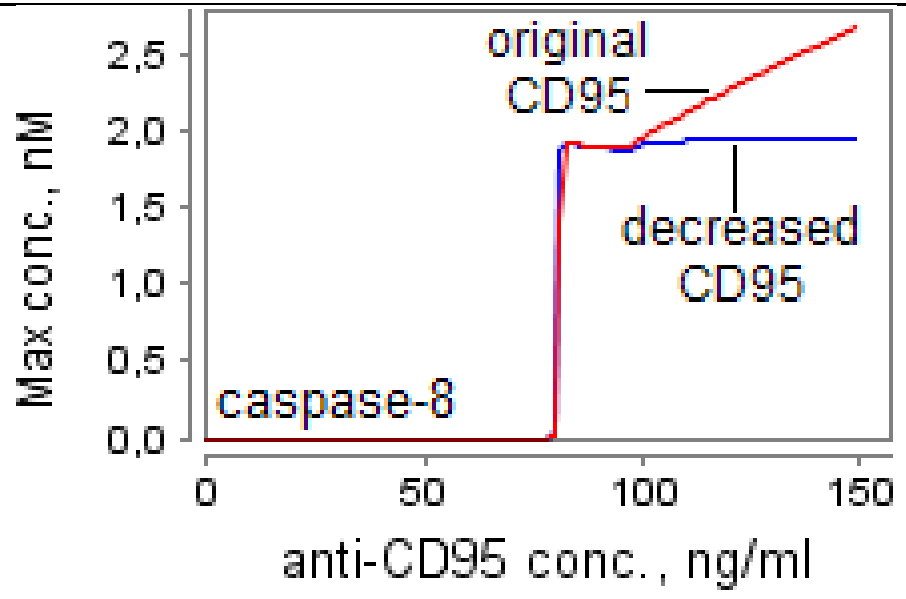 | The concentration of anti-CD95 required for the apoptosis induction (the apoptotic threshold), is within the range of 30-100 ng/ml. This range remains the same for CD95 decreased by about 12-fold.<br><br>The simulation time, which we used to reproduce this prediction, was 60 hours. |
| 2* | 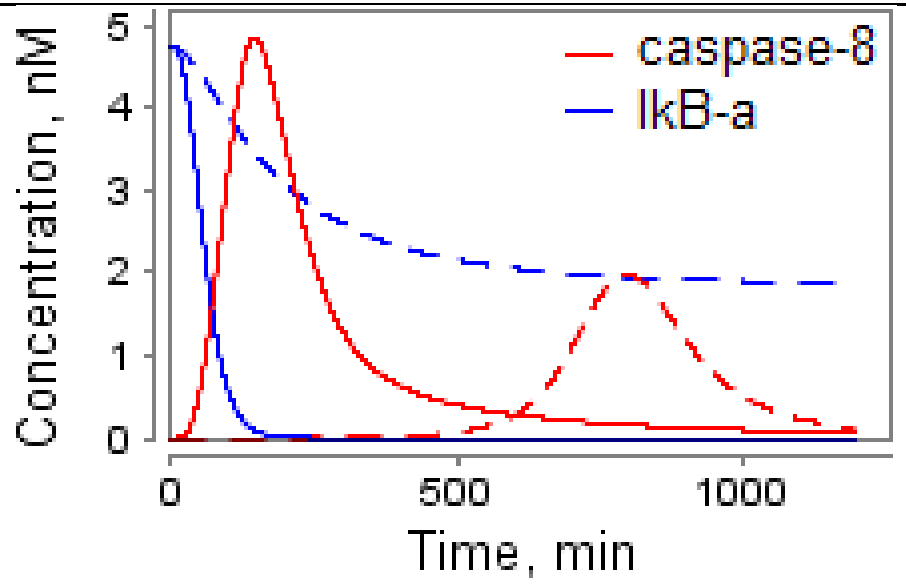 | The decreased receptor number results in impairment of both CD95- and NF-κB-signaling pathways. To test this prediction, Neumann *et al.* considered levels of caspases-8 cleavage and IκB-α degradation for the original (solid lines) amount of CD95 and the amount decreased by about 12-fold (dashed lines). The concentration of anti-CD95 was 500 ng/ml. |
| 3* | 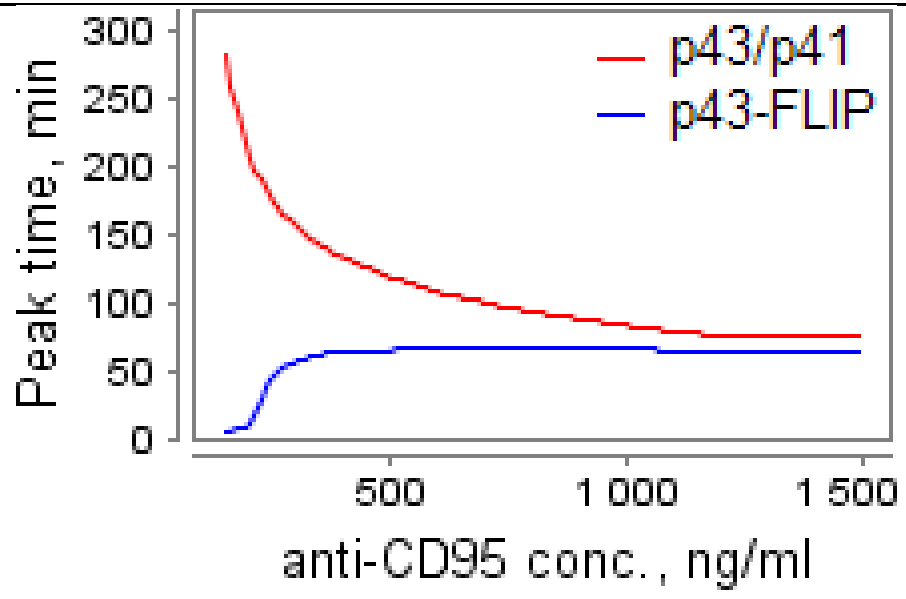 | Along with increasing the concentration of anti-CD95 from 500 ng/ml to 1500 ng/ml, p43/p41 peaks earlier, while there is almost no difference for p43-FLIP. |
| 4 | 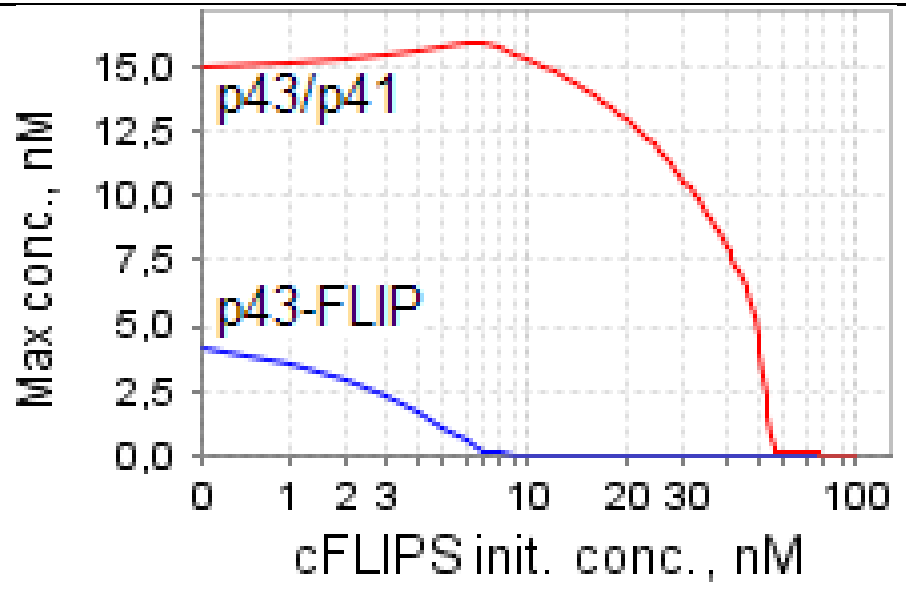 | Increased concentrations of cFLIPS inhibit both apoptotic and NF-κB pathways, although p43-FLIP generation is inhibited at a lower threshold than p43/p41 generation. |
| 5* | 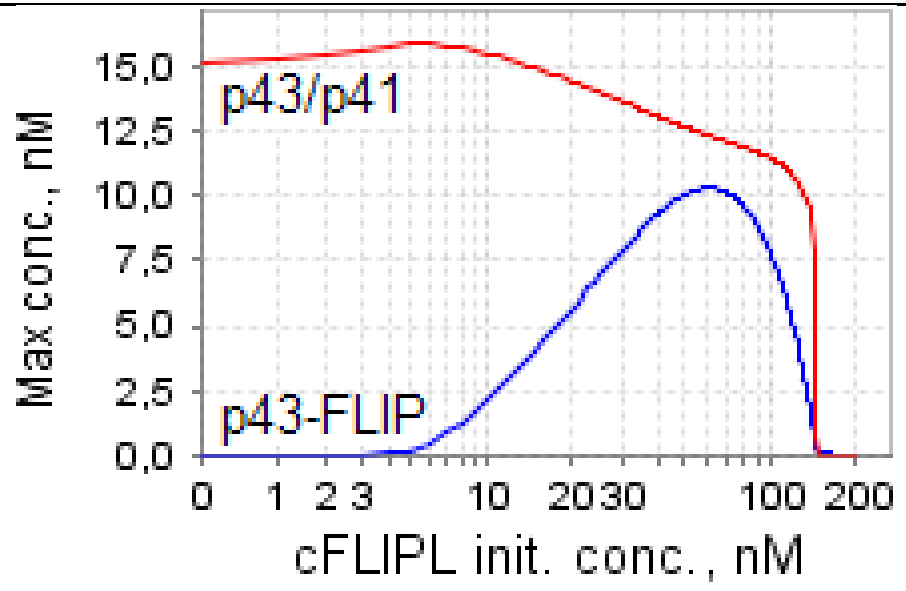 | Increasing the concentration of cFLIPL leads to a steep increase in p43-FLIP generation until it reaches a maximum, after which the curve drops. Lowered levels of cFLIPL result in very little p43-FLIP but almost unchanged levels of p43/p41.<br><br>At very high concentrations of cFLIPL no p43-FLIP is generated. This drop-off was not observed experimentally by the authors. |
| 6* | 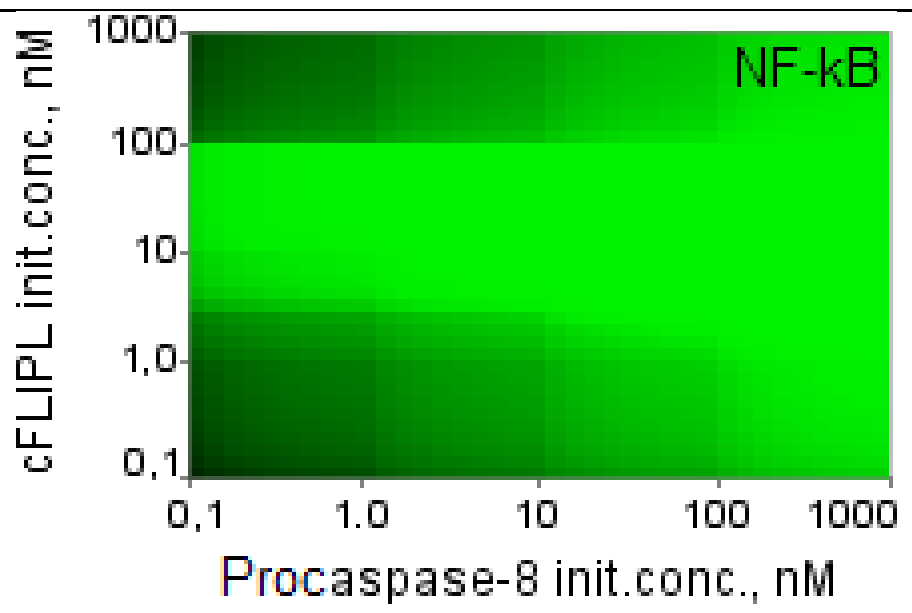 | Only an intermediate level of cFLIPL promotes NF-kB activation. Decreased levels of procaspase-8 lead to a significantly lower amount of p43-FLIP and, subsequently, NF-κB. The figures show of logarithmic dependence of the maximal NF-κB concentration on the initial values of procaspase-8 and cFLIPL |
| 7* | 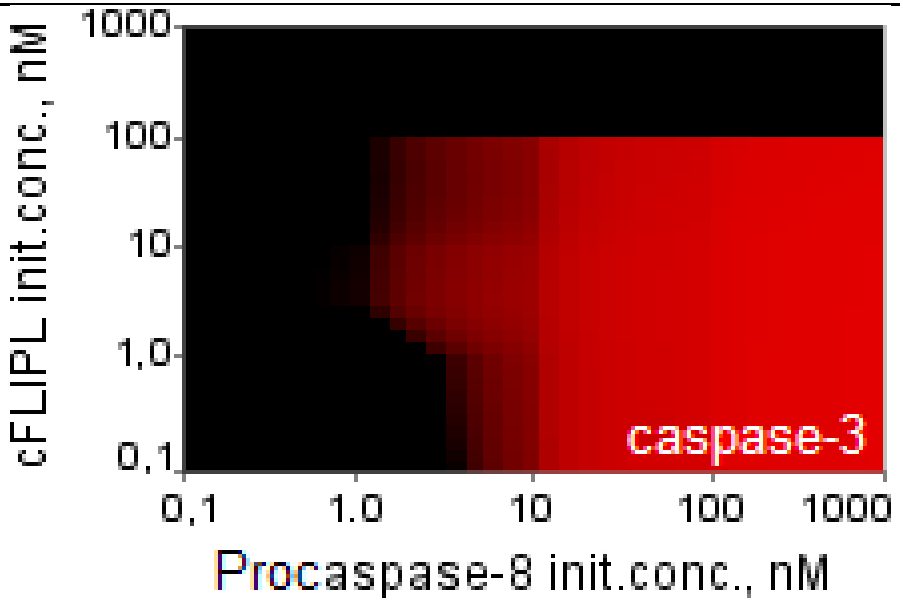 | High cFLIPL or low procaspase-8 concentrations cause suppression of apoptosis. The figures show the same dependence as considered in the previous prediction, but with caspases-3 instead of NF-κB. |

All the predictions marked with an asterisk were experimentally tested by Neumann *et al.* and confirmed, unless otherwise noted. The simulation time in predictions 3-7 was 360 min. The concentration of anti-CD95 considered by the authors in predictions 4-7 was 1000 ng/ml.

**Table 6 – Predictions of the models for SKW 6.4 cells**

| № | The models behavior | Experimental observations by Bentele *et al.* and predictions of the models |
|---|---|---|
| 1 | 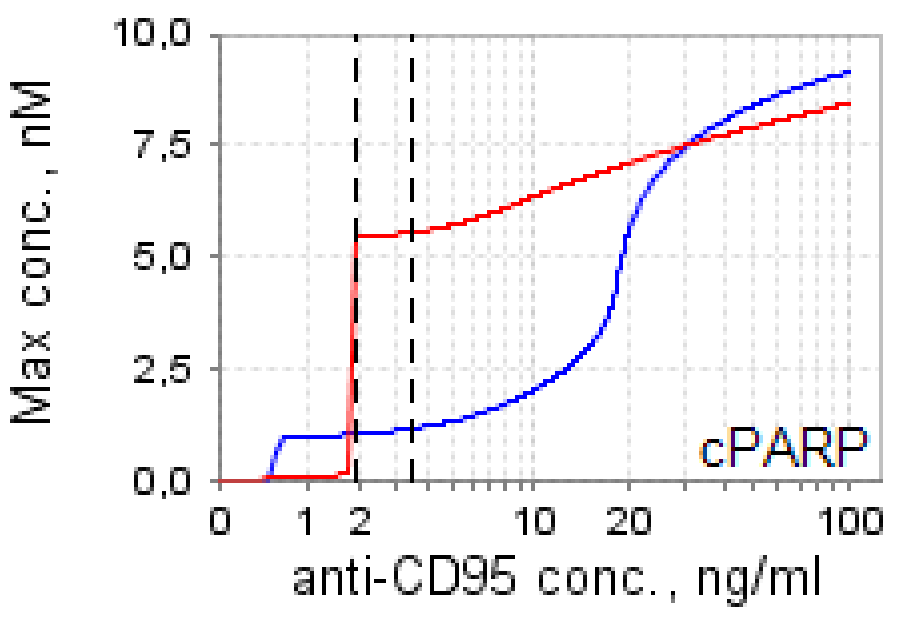 | **Experiments by Bentele *et al.*:**<br>– for 1 ng/ml of anti-CD95, PARP cleavage was not observed;<br>– the measured death rate for 10 ng/ml of anti-CD95 was 20-30%.<br>**Original model (red):**<br>– the apoptotic threshold is 1.9 ng/ml;<br>– cPARP concentration rises dramatically within an extremely narrow interval of anti-CD95 levels overcoming the apoptotic threshold.<br>**Composite model (blue):**<br>– the apoptotic threshold is 3.5 ng/ml;<br>– cPARP concentration rises in a smooth manner along with the increase of anti-CD95 level. |
| 2 | 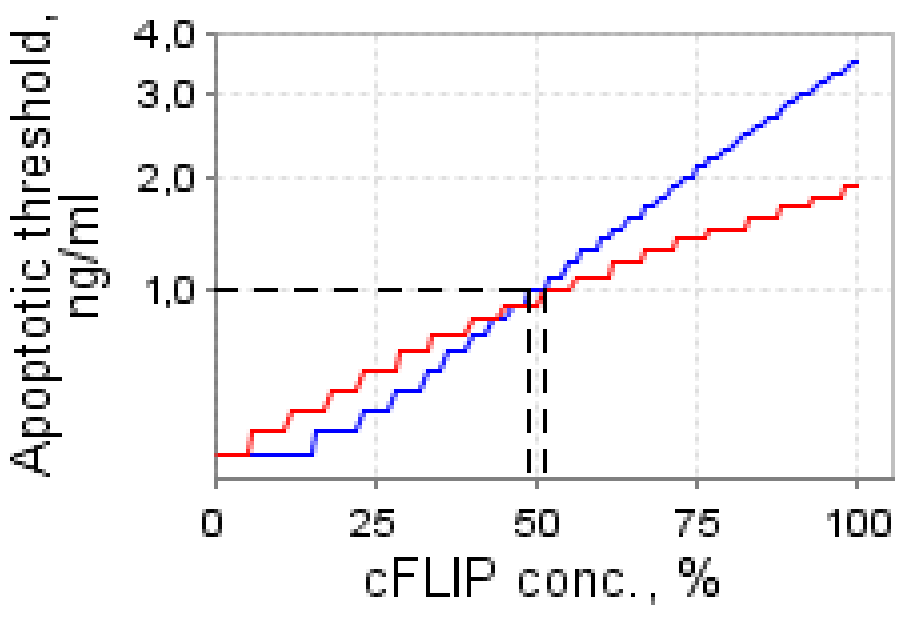 | **Experiments by Bentele *et al.*:** down-regulation of cFLIP in SKW 6.4 cells by addition of cyclohexamide resulted in cell death (40% for 1 day) already upon 1 ng/ml of anti-CD95. The level of cFLIP was decreased to 70%.<br>**Original (red) and composite (blue) models:**<br>– the apoptotic threshold is highly sensitive to the concentration of cFLIP;<br>– decreasing the initial concentration of cFLIP by more than 51% and 49% for the original and composite models, respectively, leads to cell death upon stimulation by 1 ng/ml of anti-CD95. |
| 3 | 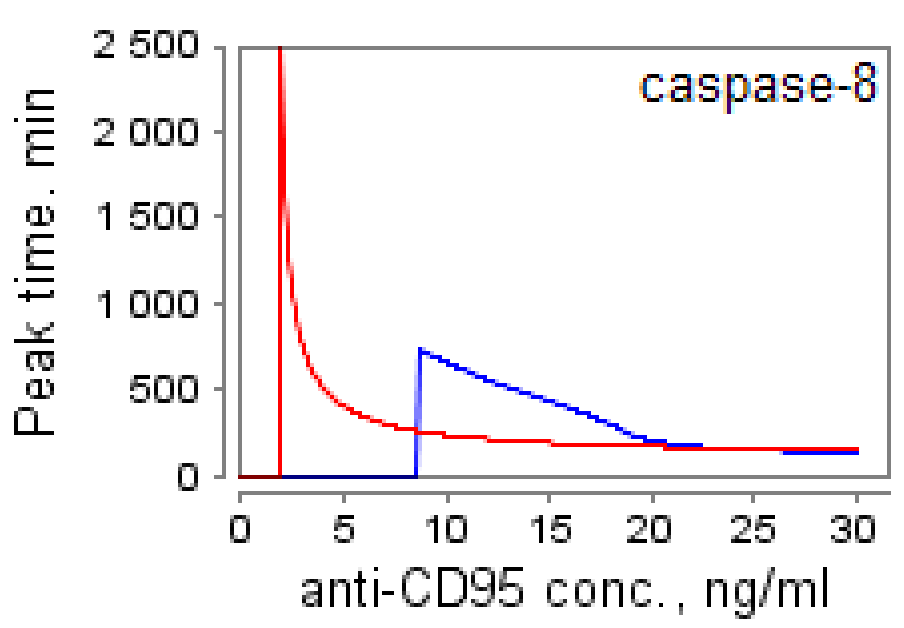 | **Experiments by Bentele *et al.*:** in the 10 ng/ml activation scenario, a significant increase of caspase-8 was observed after more than 4 hours.<br>**Original (red) and composite (blue) models:** anti-CD95 concentrations which are slightly above the apoptotic threshold result in caspase-8 activation after a delay of many hours.<br><br>The figure shows peak times of caspase-8 concentration exceeding 0.1% of the initial procaspase-8 level. |
| 4 | 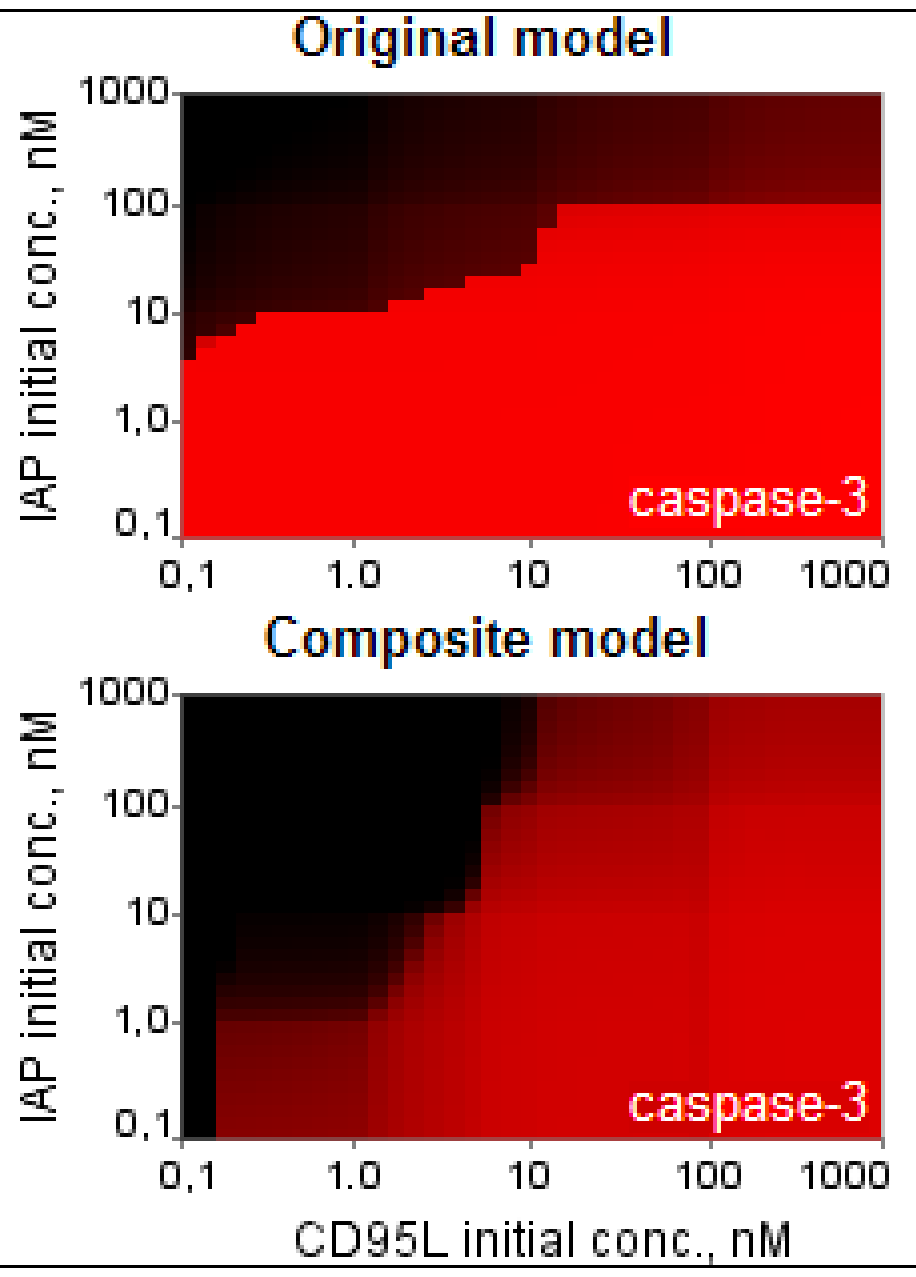 | **Original model:**<br>– low concentrations of IAP (less than 1 nM) result in complete cell death;<br>– high concentrations of IAP prevent a significant increase of caspase-3 even for high concentrations of the ligand.<br>**Composite model:**<br>– low concentrations of IAP (less than 1 nM) block apoptosis for CD95L less than 0.3 nM;<br>– high concentrations of CD95L lead to cell death.<br><br>The figures show logarithmic dependence of the maximal caspases-3 concentration on initial values of IAP and CD95L. |

The simulation time was 2880 min (2 days) in all predictions. The apoptotic threshold in the prediction 1 is the concentration of anti-CD95 after which cPARP amount exceeds 10% of the initial PARP level.

# Supplementary tables

### Table 1S – Notations of parameters used in the paper and in the original models

| Parameters notation in the paper | Parameters notation by Bentele *et al.*, 2004 |
|---|---|
| $f_{degr}$ | $f_{degrad}$ |
| $k_{Apop}$ | $k_Apoptosome$ |
| $k_d$ | $K_DEGRAD$ |
| $k_{dd}$ | $K_DEGRAD_deathSub$ |
| $k_{ds}$ | $K_DEGRAD_steady$ |
| $k_{DISC_FLIP}$ | $k_DISC_FLIP$ |
| $k_{DISC_pro8}$ | $k_DISC_procas8$ |
| $k_{DFp8}$ | $k_DISC_FLIP_to_cas8_IM$ |
| $k_{LR}$ | $k_LR$ |
| $k_{32}$ | $k_cas_3_2$ |
| $k_{36}$ | $k_cas_3_6$ |
| $k_{38}$ | $0.145 \cdot k_cas_6_8$ |
| $k_{68}$ | $k_cas_6_8$ |
| $k_{78}$ | $k_cas_8_7$ |
| $k_{83}$ | $k_cas_8_3$ |
| $k_{3act}$ | $0.18 \cdot k_cas7_apop_activity$ |
| $k_{36act}$ | $k_cas36_apop_activity$ |
| $k_{39IAP}$ | $k_cas39_IAP$ |
| $k_{7act}$ | $k_cas7_apop_activity$ |
| $k_{8Bid}$ | $k_cas8_Bid$ |
| $Km_{32}$ | $Km_cas_3_2$ |
| $Km_{36}$ | $Km_cas_3_69$ |
| $Km_{367act}$ | $Km_cas367_apop_activity$ |
| $Km_{38}$ | $Km_cas_6_8$ |
| $Km_{68}$ | $Km_cas_6_8$ |
| $Km_{78}$ | $Km_cas_8_7$ |
| $Km_{8Bid}$ | $Km_cas28_Bid$ |
| $Km_{893}$ | $Km_cas_89_3$ |
| $x_{aa}$ | $x_{apop\ activity}$ |
| | |
| **Parameters notation in the paper** | **Parameters notation by Neumann *et al.*, 2004** |
| $k_{p43-FLIP_IKK}$ | $k13$ |

### Table 2S – Table of all reactions (excepting degradation) and rate constants in the composite model

| № | Reactions | Kinetic laws | Rate constants ($nM^{-1}min^{-1}$; $min^{-1}$) | Comments[1] |
|---|---|---|---|---|
| nr1 | CD95L + CD95R:FADD → DISC | $n_{k1} \cdot C_{CD95L} \cdot C_{CD95R}$ | 1.0 | $k1$, Neumann *et al.*, 2010 |
| nr2 | pro8 + DISC → DISC:pro8 | $n_{k2} \cdot C_{DISC} \cdot C_{pro8}$ | $1.277248E-4$ | $k2$, Neumann *et al.*, 2010 |
| nr3 | DISC + FLIPL → DISC:FLIPL | $n_{k3} \cdot C_{DISC} \cdot C_{FLIPL}$ | 0.6693316 | $k3$, Neumann *et al.*, 2010 |
| nr5 | DISC:pro8 + pro8 → 2 x p43p41 | $n_{k5} \cdot C_{DISC:pro8} \cdot C_{pro8}$ | $5.946569E-4$ | $k5$, Neumann *et al.*, 2010 |
| nr6 | DISC:pro8 + FLIPL → p43-FLIP | $n_{k6} \cdot C_{DISC:pro8} \cdot C_{FLIPL}$ | 0.9999999 | $k6$, Neumann *et al.*, 2010 |
| nr7 | DISC:pro8 + FLIPS → DISC:pro8:FLIPS | $n_{k7} \cdot C_{DISC:pro8} \cdot C_{FLIPS}$ | 0.8875063 | $k7$, Neumann *et al.*, 2010 |
| nr8 | DISC:FLIPL + pro8 → p43FLIP | $n_{k5} \cdot C_{DISC:FLIPL} \cdot C_{pro8}$ | $5.946569E-4$ | $k5$, Neumann *et al.*, 2010 |
| nr9 | DISC:FLIPL + FLIPL → DISC:$FLIPL_2$ | $n_{k6} \cdot C_{DISC:FLIPL} \cdot C_{FLIPL}$ | 0.9999999 | $k6$, Neumann *et al.*, 2010 |
| nr10 | DISC:FLIPL + FLIPS → DISC:FLIPL:FLIPS | $n_{k7} \cdot C_{DISC:FLIPL} \cdot C_{FLIPS}$ | 0.8875063 | $k7$, Neumann *et al.*, 2010 |
| nr14 | 2 x p43p41 → casp8 | $n_{k8} \cdot C^2_{p43p41}$ | $8.044378E-4$ | $k8$, Neumann *et al.*, 2010 |
| nr15_m | pro3 –casp9→ casp3 | $n_{k9} \cdot C_{casp9} \cdot C_{pro3}$ | 0.04920673 | **Fitted value**, $k9 = 0.002249759$, Neumann *et al.*, 2010 |
| nr16 | pro8 -casp3→ p43p41 | $n_{k10} \cdot C_{pro8} \cdot C_{casp3}$ | 0.01205258 | **Fitted value**, $k10 = 0.1205258$, Neumann *et al.*, 2010 |
| nr19 | p43FLIP → p43-FLIP:IKK* | $n_{k13} \cdot IKK_0 \cdot C_{p43FLIP}$ | $n_{k13} = 7.20426E-4$, $IKK_0 = 5.772825$ | $k13$, $C_{IKK}(0)$, Neumann *et al.*, 2010 |
| nr20 | NF-κB:IκB -p43-FLIP:IKK→ NF-κB:IκB-P | $n_{k14} \cdot C_{\mathrm{NF\kappa B:I\kappa B}} \cdot C_{p43FLIP:IKK^*}$ | 0.3588224 | $k14$, Neumann *et al.*, 2010 |
| nr21 | NF-κB:IκB-P → NF-κB* | $n_{k15} \cdot C_{\mathrm{NF\kappa B:I\kappa BP}}$ | 3.684162 | $k15$, Neumann *et al.*, 2010 |
| br6*_m | pro9 -tBid→ casp9 | $b_{k6} \cdot C_{tBid} \cdot C_{pro9}$ | 0.06310456 | – |
| br7* | Bid -casp8→ tBid | $b_{k7} \cdot C_{casp8} \cdot C_{Bid}$ | $6.0004E-4$ | **Fitted value**, $k_{8Bid}/Km_{8Bid} = 5.3325E-4$, the reduced Bentele's model |
| br8* | pro2 -casp3→ (cleavage) | $\frac{b_{k8} \cdot C_{casp3} \cdot C_{pro2}}{b_{Km8} + C_{pro2}}$ | $b_{k8} = 15.63123994$, $b_{Km8} = 55.57400642$ | **Fitted values**, $k_{32} = 0.18137$, $Km_{32} = 55.574$, Bentele *et al.*, 2004 |
| br9* | pro3 -casp8→ casp3 | $\frac{b_{k9} \cdot C_{casp8} \cdot C_{pro3}}{b_{Km9} + C_{pro3}}$ | $b_{k9} = 0.0$ (HeLa) or 0.27246 (SKW 6.4), $b_{Km9} = 1.03542544$ | **Fitted values**, $k_{83} = 1.90358$, $Km_{893} = 89.02911547$, Bentele *et al.*, 2004 |
| br10* | pro7 -casp8→ (cleavage) | $b_{k10} \cdot C_{casp8} \cdot C_{pro7}$ | 10.89517864 | **Fitted value**, $k_{78}/Km_{78} = 0.01016$, the reduced Bentele's model |
| br11* | PARP –casp3→ cPARP | $b_{k11} \cdot C_{casp7} \cdot C_{PARP}$ | 73.49560485 | **Fitted value**, $0.18 \cdot k_{7act}/Km_{367act} = 0.00824$, the reduced Bentele's model |
| br12* | casp3 + IAP → (inhibition) | $b_{k12} \cdot C_{casp3} \cdot C_{IAP}$ | 0.01 | **Fitted value**, $k_{39IAP} = 45824.409 \cdot 5 \cdot 10^{-5}$, Bentele *et al.*, 2004 |
| br13* | -casp3→ Apoptotic activity | $b_{k13} \cdot (1.0 - x_{aa}) \cdot C_{casp3}$ | 5956.81008868 | **Fitted value**, $(k_{36act} + 0.18 \cdot k_{7act})/Km_{367act} = 0.01053$, Bentele et al., 2004 |

[1] The column contains basic parameters or expressions (and their values if they were refitted) in the combined models.

### Table 3S – Table of degraded species and kinetic laws of degradation in the composite model

| Degraded species | Degradation rates | Rate constants ($min^{-1}$) |
|---|---|---|
| pro2, pro3, pro8, DISC:pro8, DISC:pro8:FLIPS, pro9, Bid, PARP | $f_{degr} = degr \cdot x_{aa}^2$ | $degr = \begin{cases} 0.0, & \text{HeLa cells,} \\ 0.0542, & \text{SKW 6.4 cells,} \quad C_{CD95L} = 5\,\mu g/ml, \\ 0.0084, & \text{SKW 6.4 cells,} \quad C_{CD95L} = 200\,ng/ml, \\ 0.0028, & \text{SKW 6.4 cells,} \quad 1\ ng/ml \le C_{CD95L} \le 100\,ng/ml. \end{cases}$ |
| casp3 | $f_{degr} = degr \cdot x_{aa}^2 + n_{k12}$ | $n_{k12} = 0.1502914$, ($k12$, Neumann *et al.*, 2010) |
| casp8 | $f_{degr} = degr \cdot x_{aa}^2 + n_{k11}$ | $n_{k11} = 0.02891451$, ($k11$, Neumann *et al.*, 2010) |
| casp9 | $f_{degr} = degr \cdot x_{aa}^2 + degr_{casp9}$ | $degr_{casp9} = 0.10025628$ |
| cPARP | $f_{degr} = degr_{cPARP}$ | 0.01748, (K_DEGRAD_PARP, Bentele *et al*., 2004) |
| tBid | $f_{degr} = degr_{tBid}$ | 0.04999612 |
| p43-FLIP:IKK* | $f_{degr} = n_{k16}$ | 0.02229912, ($k16$, Neumann *et al.*, 2010) |
| Apoptotic activity ($x_{aa}$) | $degr_{apop\ activity}$ | 0.00219407 |

### Table 4S – Table of nonzero initial concentrations in the composite model

| Species | Initial concentrations (nM) | | | |
|---|---|---|---|---|
| | **Neumann *et al*., 2010 HeLa cells** | **Bentele *et al*., 2004 SKW 6.4 cells** | **Composite model HeLa cells** | **Composite model SKW 6.4 cells** |
| CD95L | 113.22, 37.74, 18.87 | 1990.0, 79.6 | 113.22, 37.74, 18.87 | 1990.0, 79.6 |
| CD95R | – | 442.820768294033 | – | – |
| CD95R:FADD | 91.26592 | 0.0 | 91.26592 | 611.6891578799691 |
| procaspase-2 | – | 157.644193512676 | 157.644193512676 | 157.644193512676 |
| procaspase-3 | 1.443404 | 112.45433410827 | 14.43404 | 2.3438636537313413 |
| procaspase-7 | – | 18.7933134063988 | 18.7933134063988 | 18.7933134063988 |
| procaspase-8 | 64.47652 | 442.820768294033 | 64.47652 | 350.0248656584318 |
| procaspase-9 | – | 245.101295250747 | 2.9090736162783806 | 245.101295250747 |
| FLIPL | 7.398562 | 65.0213661020702 | 7.398562 | 7.398562 |
| FLIPS | 5.083923 | 65.0213661020702 | 5.083923 | 70.44906883596883 |
| IAP | – | 12.2160965349275 | 1.221610 | 1.221610 |
| Bid | – | 231.760433964353 | 5.003142624870996 | 231.760433964353 |
| IKK | 5.772825 | – | 5.772825 | 5.772825 |
| NF-κB:IκB | 4.739546 | – | 4.739546 | 4.739546 |
| PARP | – | 11.1615188752353 | 11.1615188752353 | 11.1615188752353 |

### Table 5S – Steady state analysis of the Bentele's model and the composite model

| Species | Steady state values, anti-CD95 = 200 ng/ml | | | Steady state values, anti-CD95 = 5 μg/ml | | |
|---|---|---|---|---|---|---|
| | Original model | Reduced model | Composite model | Original model | Reduced model | Composite model |
| procaspase-2 | 0.0000374 | 0.0000316 | 2.4468579 | 0.0013440 | 0.0013602 | 0.0000096 |
| procaspase-3 | – | – | – | 0.0 | 0.0 | 0.0 |
| procaspase-7 | – | – | – | 0.0 | 0.0 | 0.0 |
| procaspase-8 | 25.1700455 | 27.2240392 | 14.9514288 (total value) | 0.0000016 | 0.0 | 0.0000367 (total value) |
| procaspase-9 | 0.0 | 0.0 | 0.0 | – | – | – |
| caspase-8 | 0.0 | 0.0 | 0.0 | 0.0 | 0.0 | 0.0 |
| p43/p41 | 0.0 | 0.0 | 0.0001051 | – | – | – |
| Bid | – | – | – | 0.0411840 | 0.0342991 | 0.0000298 |
| tBid | – | – | – | 0.0901602 | 0.1040935 | 0.0 |
| PARP | 0.0 | 0.0 | 0.0 | 0.0 | 0.0 | 0.0 |
| cPARP | – | – | – | 0.0 | 0.0 | 0.0 |

### Table 6S – Steady state analysis of the Neumann's model and the composite model

| Species | Steady state values, anti-CD95 = 1500 ng/ml | | | Steady state values, anti-CD95 = 500 ng/ml | | | Steady state values, anti-CD95 = 250 ng/ml | | |
|---|---|---|---|---|---|---|---|---|---|
| | Original model | Reduced model | Composite model | Original model | Reduced model | Composite model | Original model | Reduced model | Composite model |
| procaspase-8 (total) | 12.0752771 | 12.0751903 | 12.0181928 | 2.6562426 | 2.6562649 | 2.3208092 | 9.6046242 | 9.6052138 | 13.2926881 |
| p43/p41 | 0.0010204 | 0.0010204 | 0.0000891 | 0.0009555 | 0.0009555 | 0.0000948 | 0.0008922 | 0.0008922 | 0.0000883 |
| caspase-8 | 0.0 | 0.0 | 0.0 | 0.0 | 0.0 | 0.0 | 0.0 | 0.0 | 0.0 |
| procaspase-3 | 0.1992635 | 0.1992525 | 3.8521010 | 0.1350020 | 0.1349977 | 3.7249153 | 0.1721551 | 0.1721569 | 3.8516549 |
| caspase-3 | 0.0 | 0.0 | 0.0 | 0.0 | 0.0 | 0.0 | 0.0 | 0.0 | 0.0 |
| IκB | 0.0 | 0.0 | 0.0 | 0.0000020 | 0.0000020 | 0.0000020 | 0.1519007 | 0.1526973 | 0.1526973 |
| IκB-P | 0.0 | 0.0 | 0.0 | 0.0 | 0.0 | 0.0 | 0.0 | 0.0 | 0.0 |
| p43-FLIP | 0.0 | 0.0 | 0.0 | 0.0 | 0.0 | 0.0 | 0.0 | 0.0 | 0.0 |

### Table 7S – Calculation of the mean sensitivity for the investigated apoptosis models

| Cell lines, anti-CD95 concentrations | The model by Bentele *et al.* | | | The model by Neumann *et al.* | | | The composite model, all parameters |
|---|---|---|---|---|---|---|---|
| | Original model | | Reduced model | Original model | | Reduced model | |
| | All parameters | Retained parameters | | All parameters | Retained parameters | | |
| SKW 6.4, 5 μg/ml | −34.49 | −33.79 | −29.72 | – | – | – | −146.84 |
| SKW 6.4, 200 ng/ml | −25.64 | −24.95 | −38.97 | – | – | – | −114.13 |
| HeLa, 1500 ng/ml | – | – | – | −30.71 | −30.42 | −30.29 | −30.21 |
| HeLa, 500 ng/ml | – | – | – | −79.85 | −79.52 | −78.94 | −79.88 |
| HeLa, 250 ng/ml | – | – | – | −88.05 | −87.60 | −88.70 | −97.69 |

## Table 8S – Analysis of predictions regarding apoptosis in HeLa cells as formulated by Neumann *et al*.

| №[2] | Behavior of the model by Neumann *et al.* | Behavior of the reduced model | The composite model behavior | Predictions by Neumann *et al.*[3] |
|---|---|---|---|---|
| 1* | 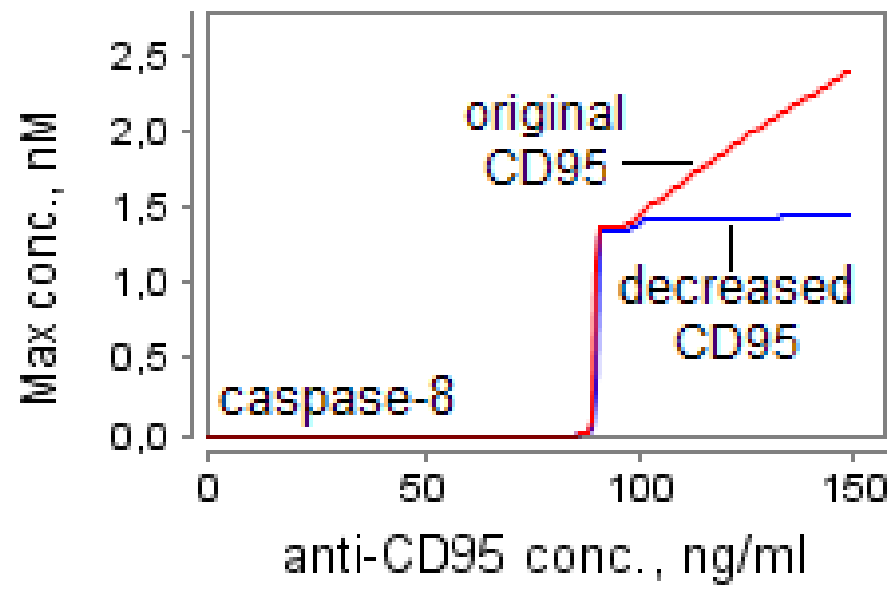 | 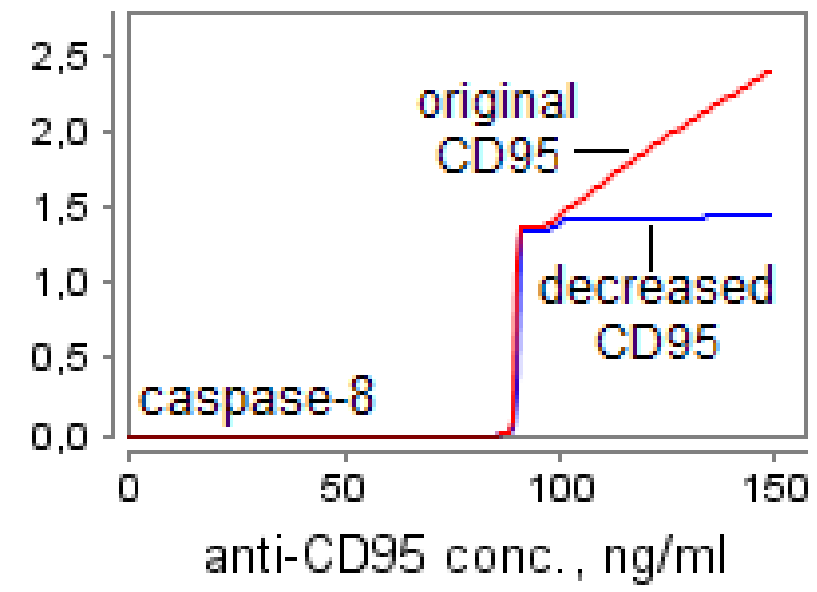 | 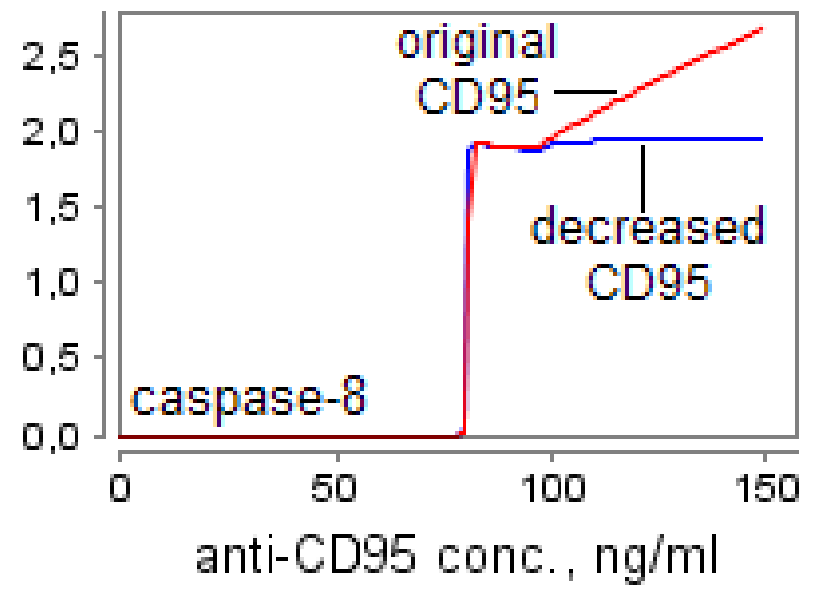 | The concentration of anti-CD95 required for the apoptosis induction (the apoptotic threshold), is within the range of 30-100 ng/ml. This range remains the same for CD95 decreased by about 12-fold.<br><br>The simulation time, which we used to reproduce this prediction, was 60 hours. |
| 2* | 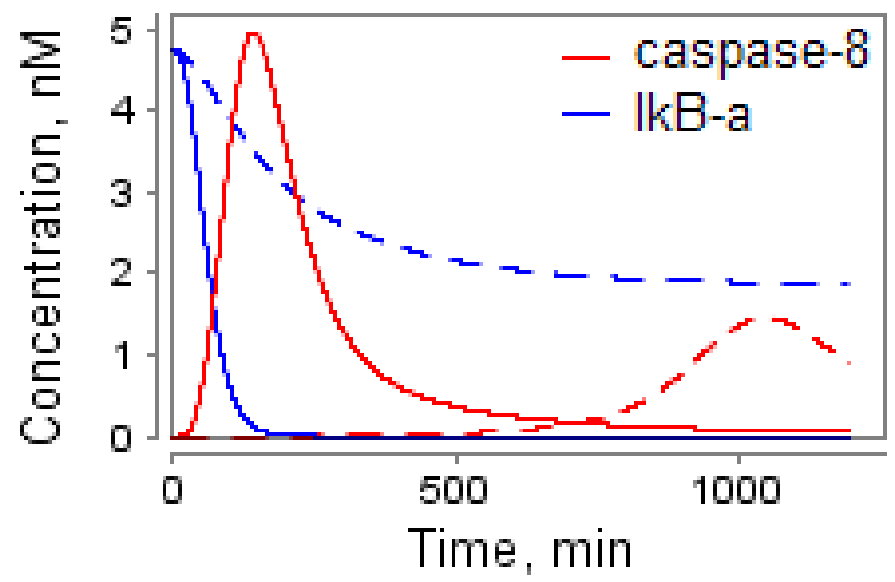 | 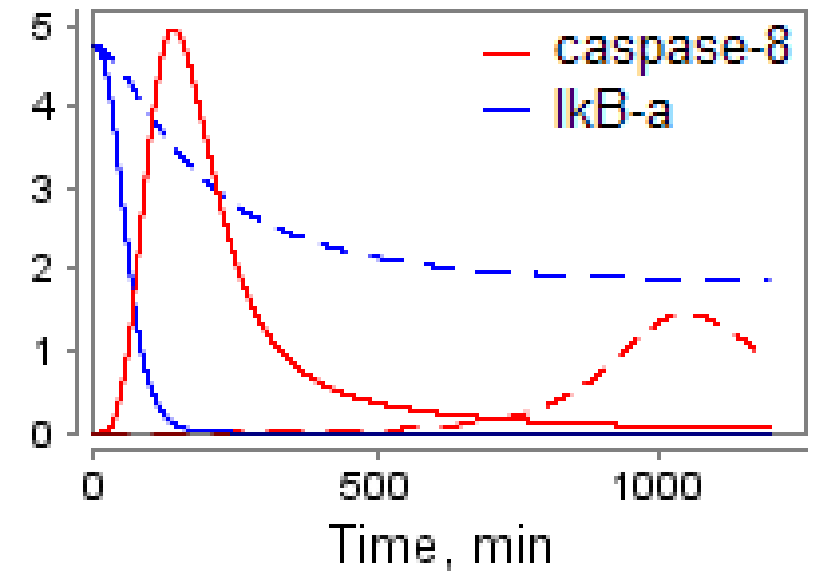 | 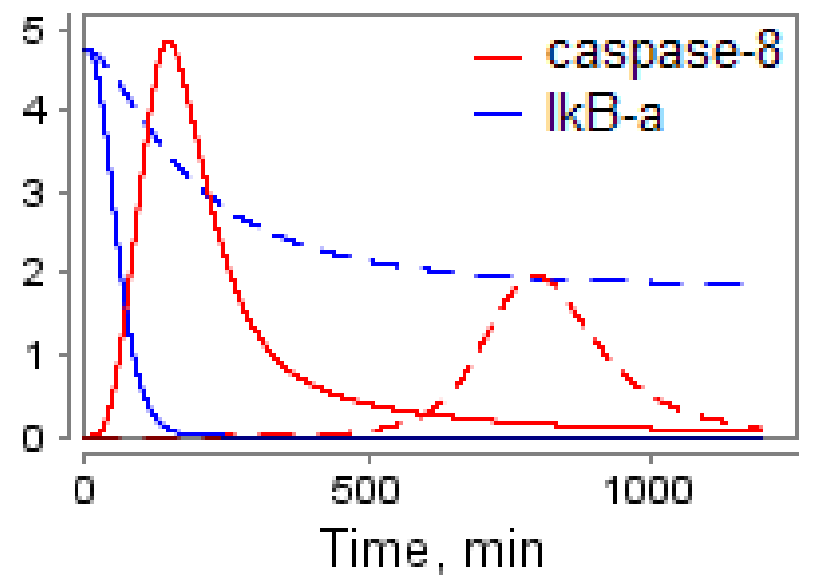 | The decreased receptor number results in impairment of both CD95- and NF-κB-signaling pathways. To test this prediction, Neumann *et al.* considered levels of caspases-8 cleavage and IκB-α degradation for the original (solid lines) amount of CD95 and the amount decreased by about 12-fold (dashed lines). The concentration of anti-CD95 was 500 ng/ml. |
| 3* | 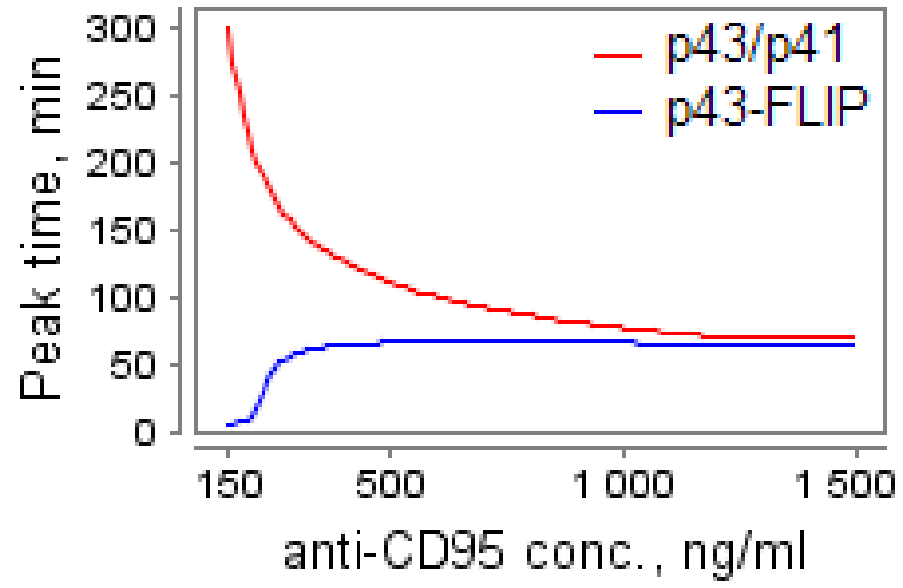 | 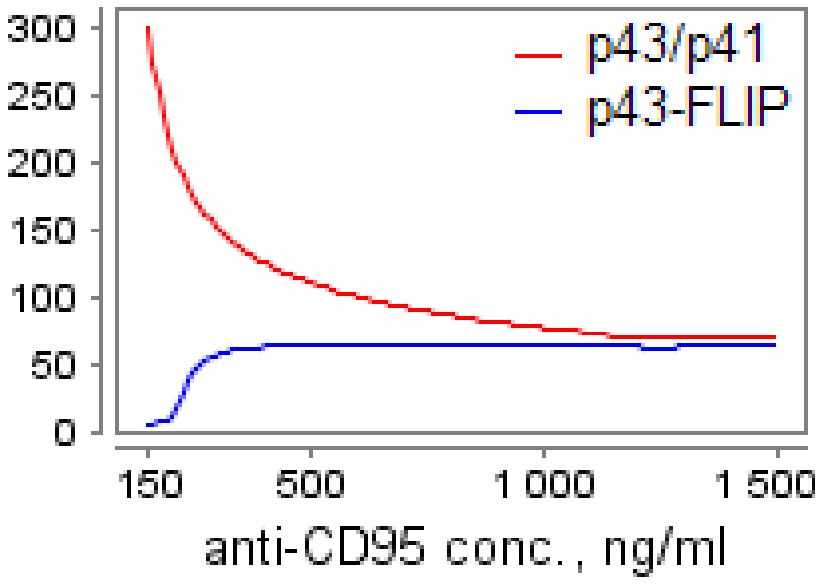 | 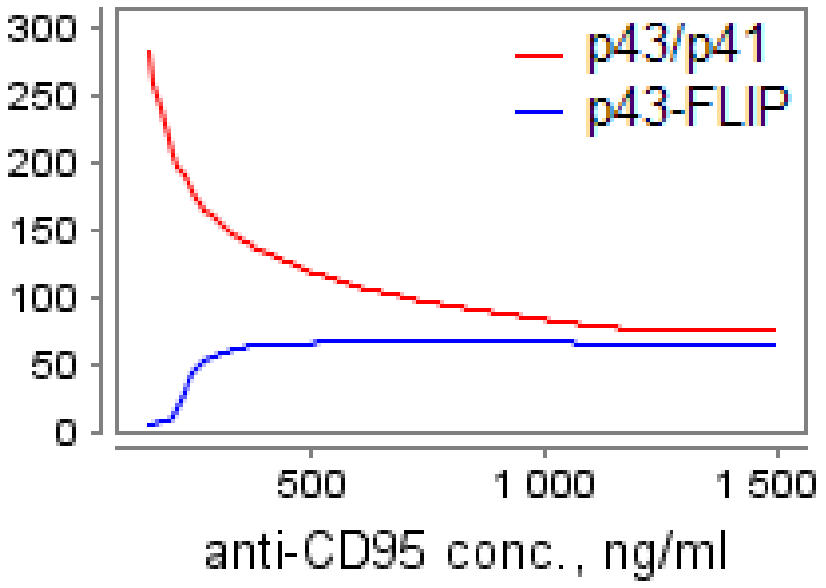 | Along with increasing the concentration of anti-CD95 from 500 ng/ml to 1500 ng/ml, p43/p41 peaks earlier, while there is almost no difference for p43-FLIP. |
| 4 | 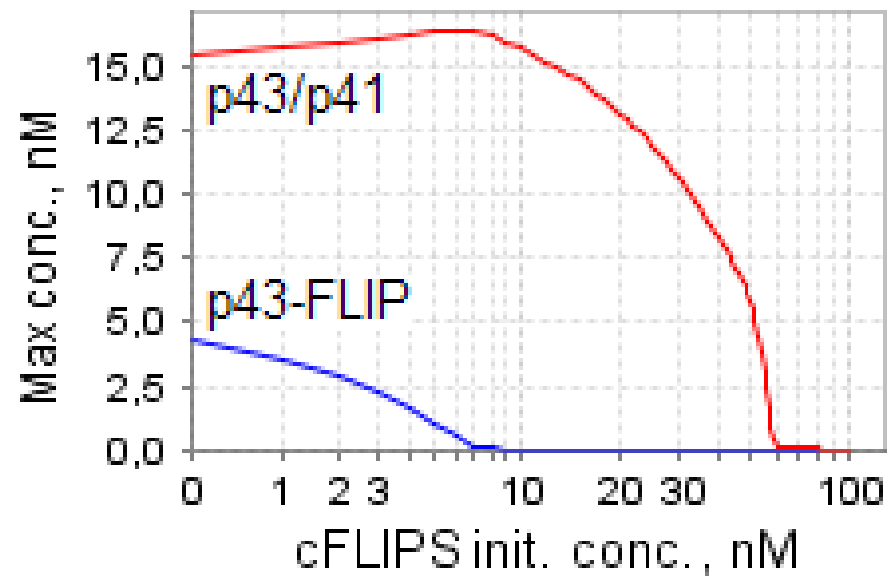 | 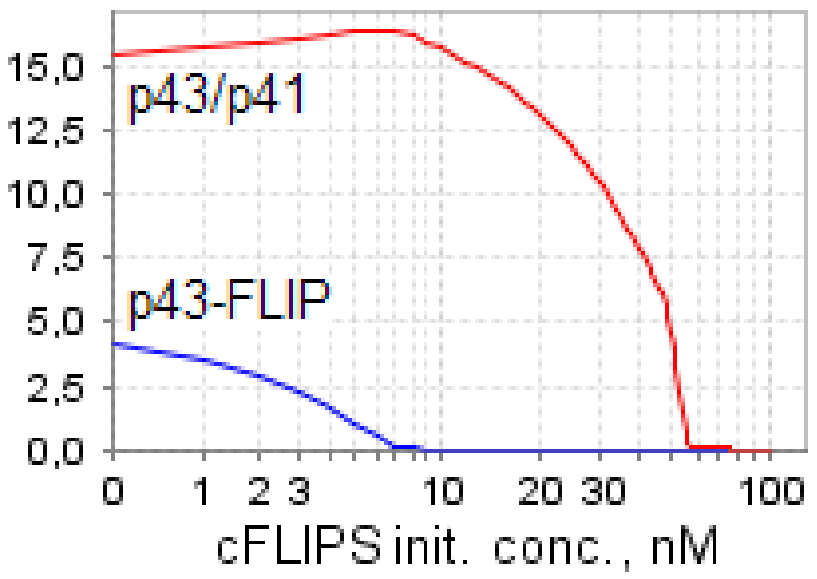 | 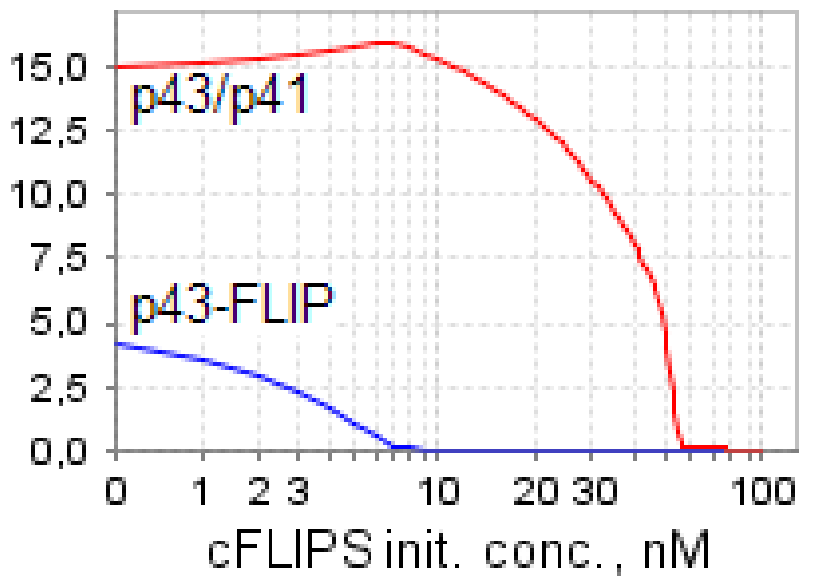 | Increased concentrations of cFLIPS inhibit both apoptotic and NF-κB pathways, although p43-FLIP generation is inhibited at a lower threshold than p43/p41 generation. |

[2] All the predictions marked with an asterisk were experimentally tested by Neumann *et al.* and confirmed, unless otherwise noted.
[3] The The simulation time in predictions 3-7 was 360 min. The concentration of anti-CD95 considered by the authors in predictions 4-7 was 1000 ng/ml.

**Table 8S (continuation)**

| № | Behavior of the model by Neumann *et al.* | Behavior of the reduced model | The composite model behavior | Predictions by Neumann *et al.* |
|---|---|---|---|---|
| 5* | 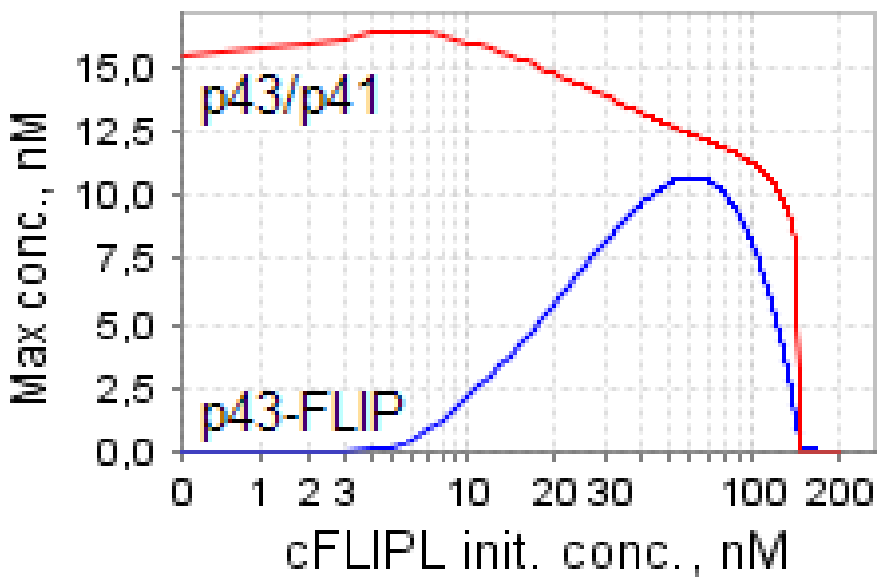 | 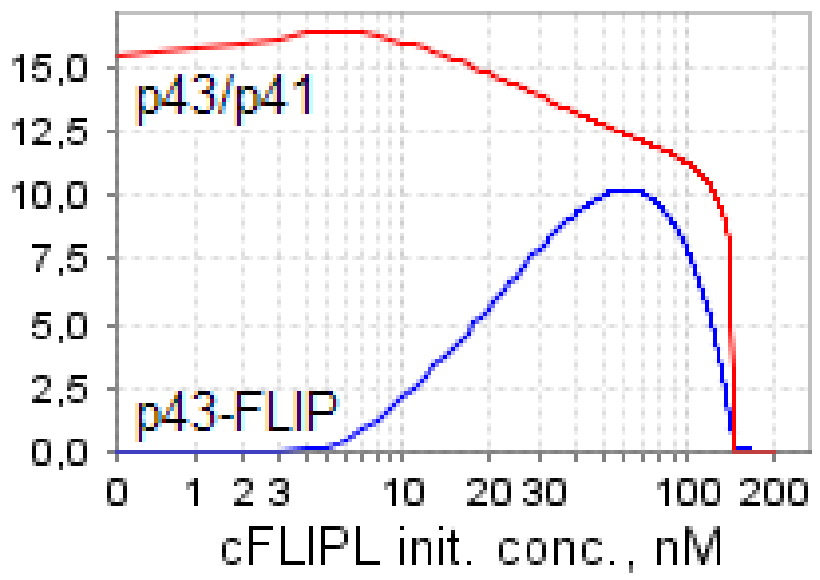 | 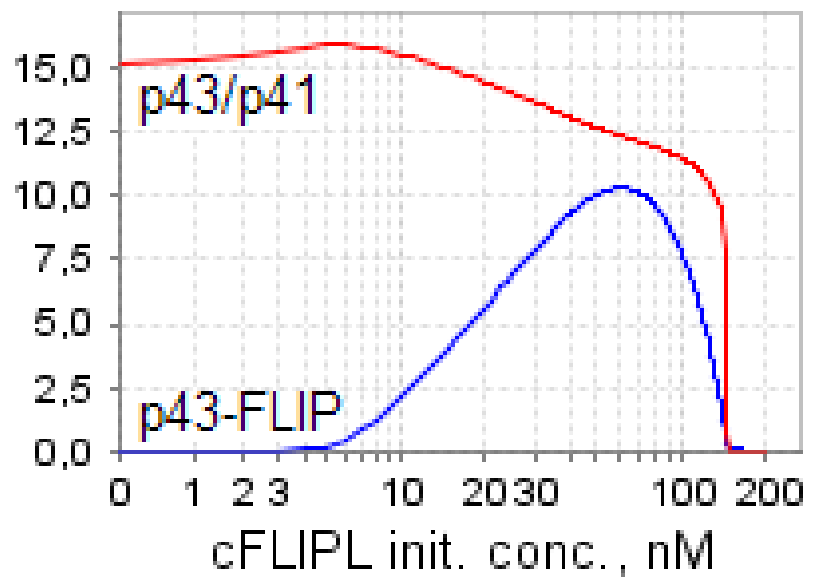 | Increasing the concentration of cFLIPL leads to a steep increase in p43-FLIP generation until it reaches a maximum, after which the curve drops. Lowered levels of cFLIPL result in very little p43-FLIP but almost unchanged levels of p43/p41.<br><br>At very high concentrations of cFLIPL no p43-FLIP is generated. This drop-off was not observed experimentally by the authors. |
| 6* | 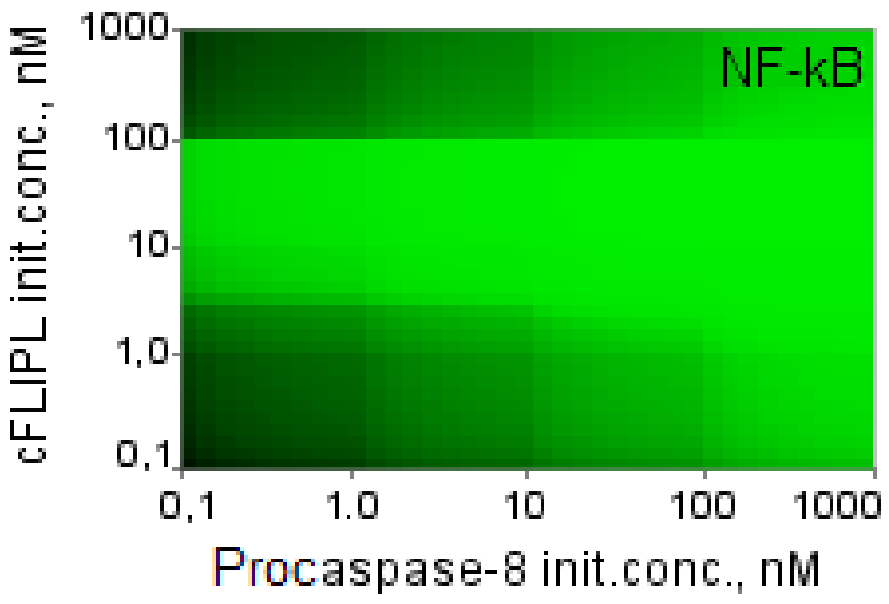 | 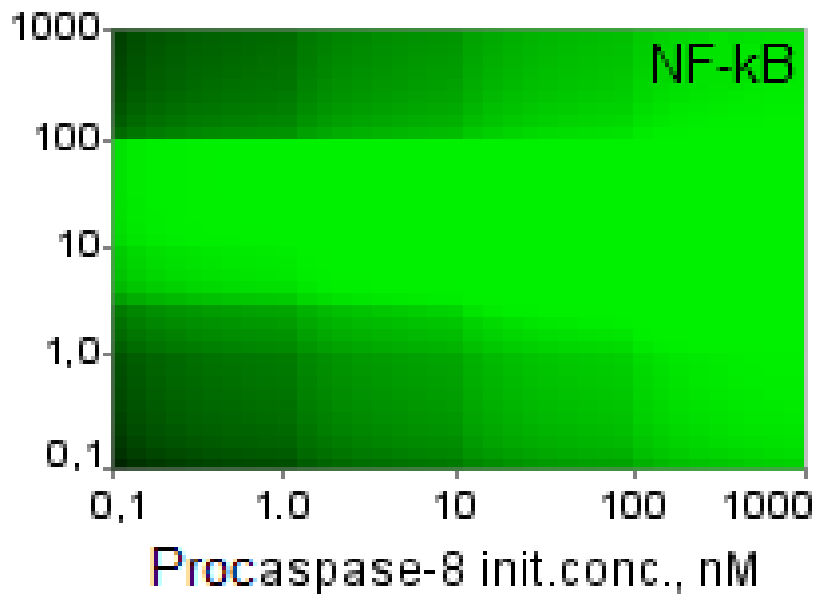 | 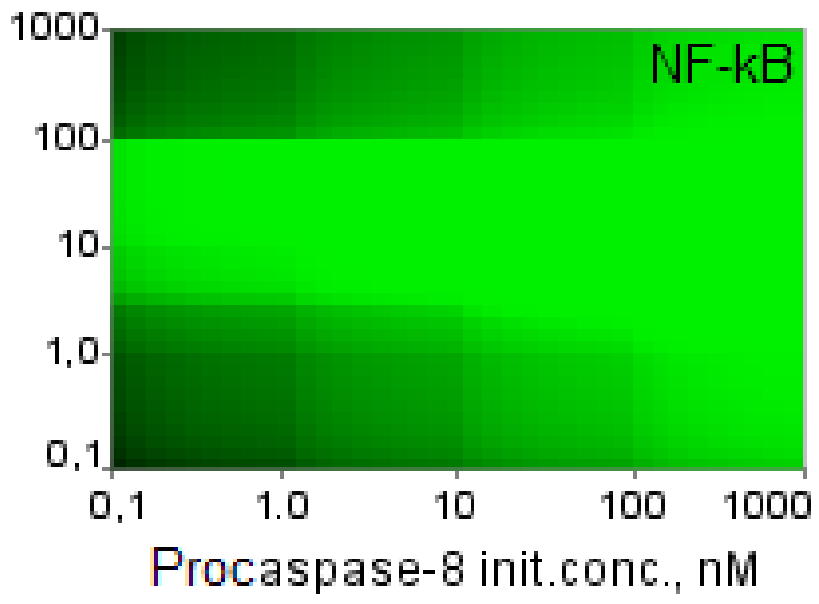 | Only an intermediate level of cFLIPL promotes NF-kB activation. Decreased levels of procaspase-8 lead to a significantly lower amount of p43-FLIP and, subsequently, NF-κB. The figures show of logarithmic dependence of the maximal NF-κB concentration on the initial values of procaspase-8 and cFLIPL |
| 7* | 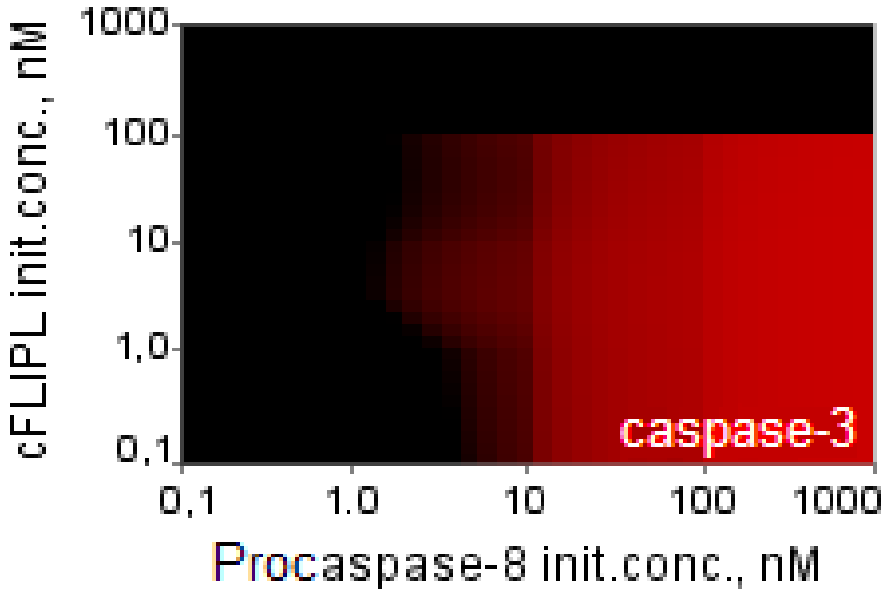 | 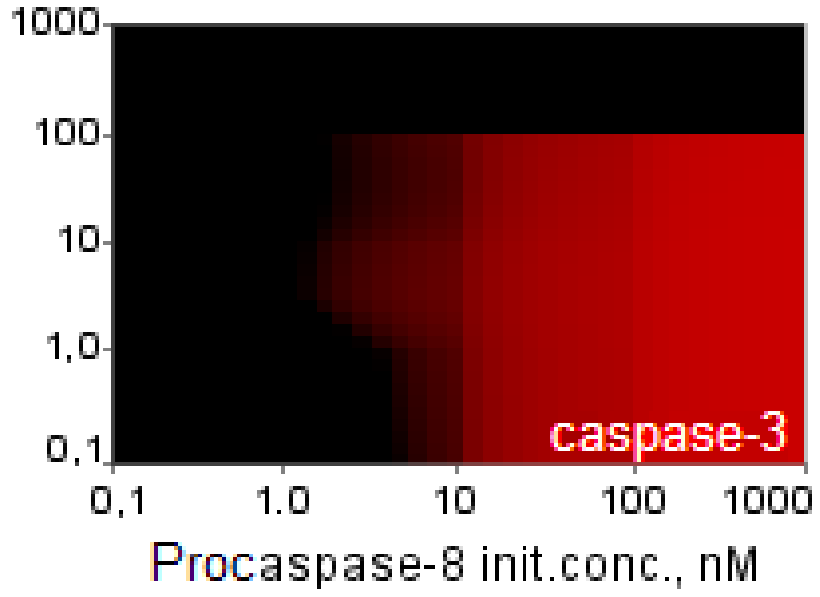 | 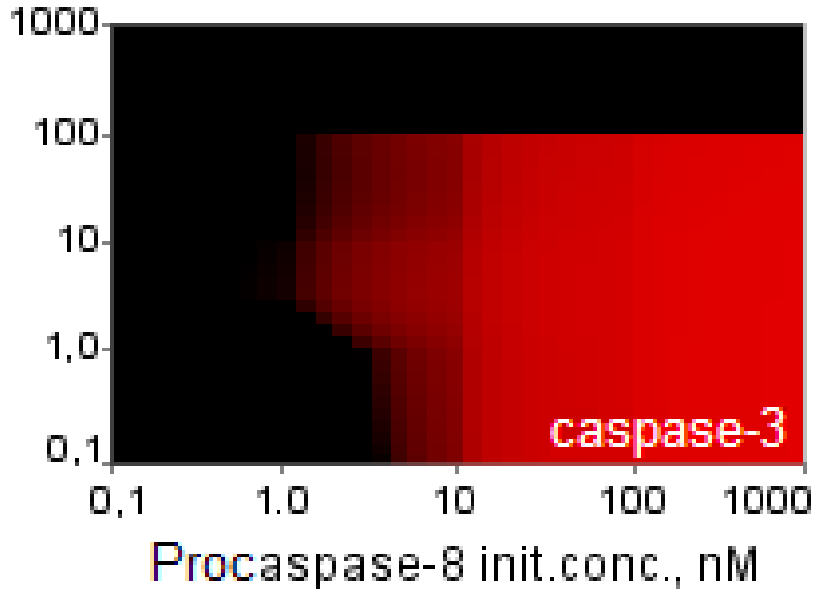 | High cFLIPL or low procaspase-8 concentrations cause suppression of apoptosis. The figures show the same dependence as considered in the previous prediction, but with caspases-3 instead of NF-κB. |

## Table 9S – Predictions of the models for SKW 6.4 cells

| № | Behavior of the model by Bentele *et al.* | Behavior of the reduced model | The composite model behavior | Experimental observations by Bentele *et al.* and predictions of the models[4] |
|---|---|---|---|---|
| 1 | 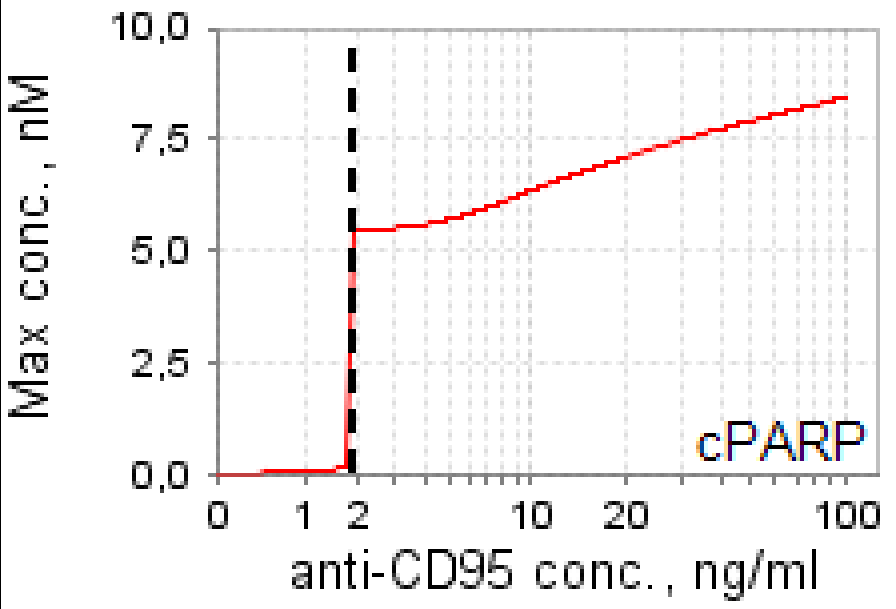 | 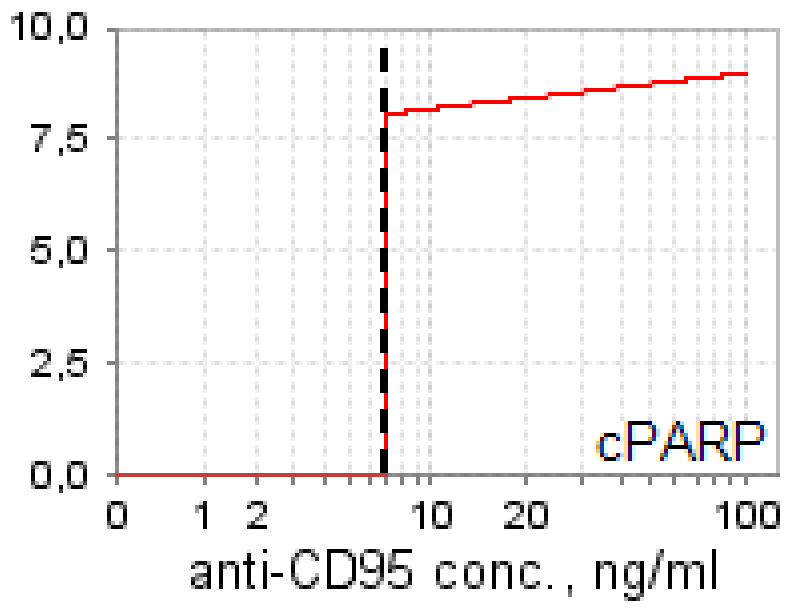 | 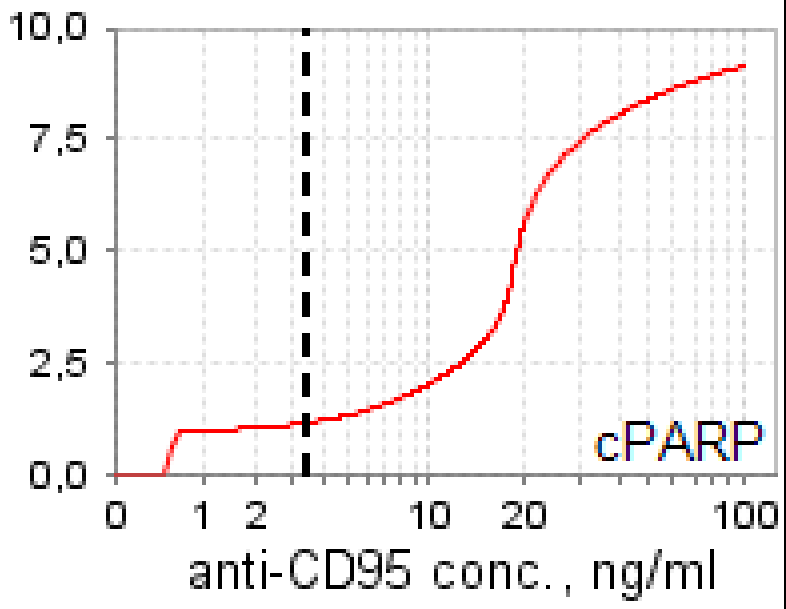 | **Experiments by Bentele *et al.*:**<br>– for 1 ng/ml of anti-CD95, PARP cleavage was not observed;<br>– the measured death rate for 10 ng/ml of anti-CD95 was 20-30%.<br>**Original and reduced models:**<br>– the apoptotic threshold[5] is 1.9 and 6.9 ng/ml for the original and reduced models, respectively;<br>– cPARP concentration rises dramatically within an extremely narrow interval of anti-CD95 levels overcoming the apoptotic threshold.<br>**Composite model:**<br>– the apoptotic threshold is 3.5 ng/ml;<br>– cPARP concentration rises in a smooth manner along with the increase of anti-CD95 level. |
| 2 | 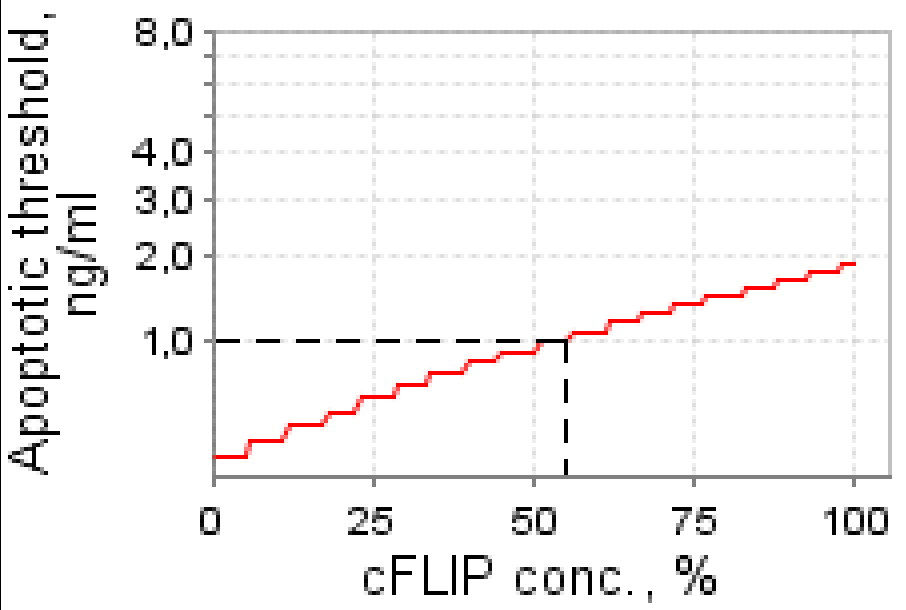 | 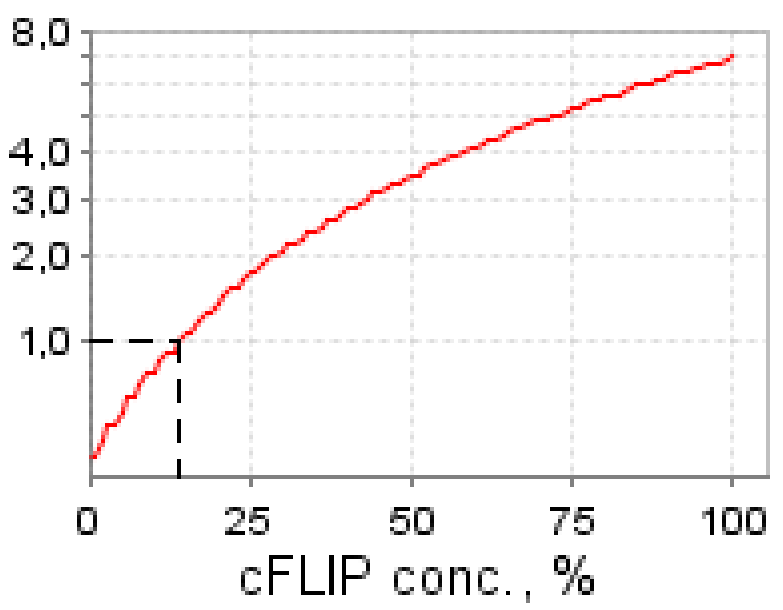 | 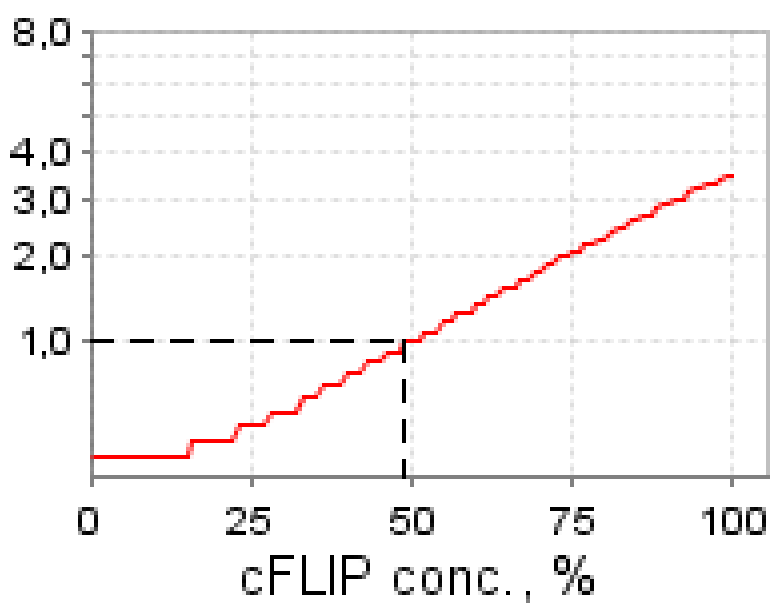 | **Experiments by Bentele *et al.*:** down-regulation of cFLIP in SKW 6.4 cells by addition of cyclohexamide resulted in cell death (40% for 1 day) already upon 1 ng/ml of anti-CD95. The level of cFLIP was decreased to 70%.<br>**All models:**<br>– the apoptotic threshold is highly sensitive to the concentration of cFLIP;<br>– decreasing the initial concentration of cFLIP by more than 51%, 86% and 49% for the original, reduced and composite models, respectively, leads to cell death upon stimulation by 1 ng/ml of anti-CD95. |
| 3 | 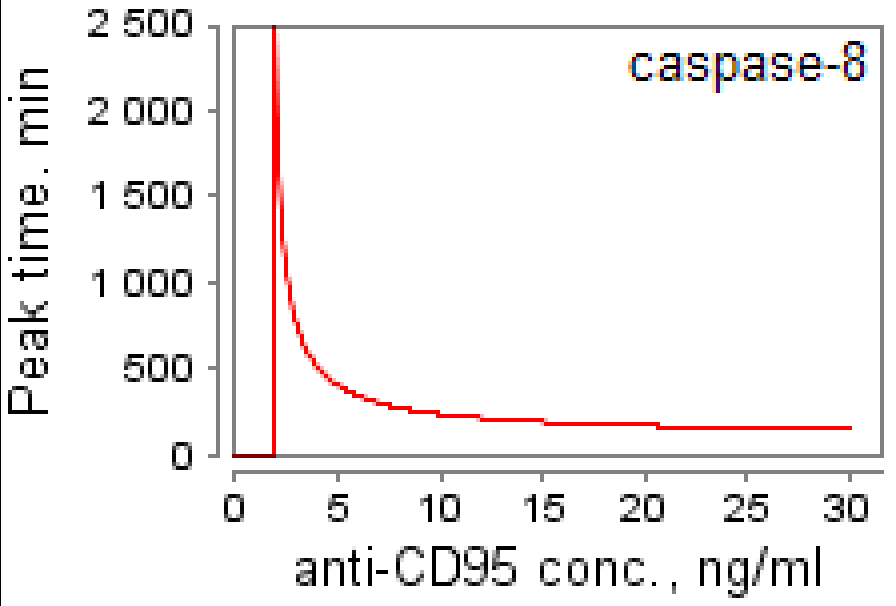 | 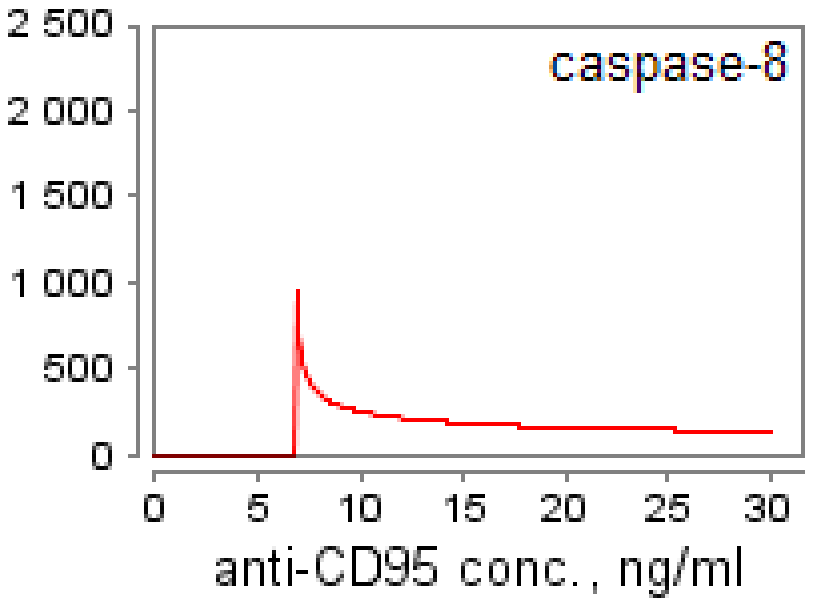 | 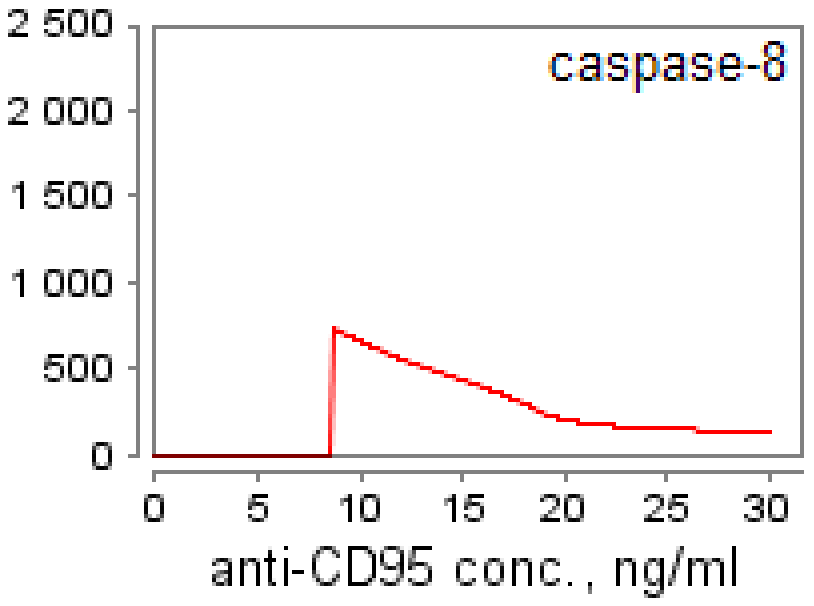 | **Experiments by Bentele *et al.*:** in the 10 ng/ml activation scenario, a significant increase of caspase-8 was observed after more than 4 hours.<br>**All models:** anti-CD95 concentrations which are slightly above the apoptotic threshold result in caspase-8 activation after a delay of many hours.<br><br>The figures show peak times of caspase-8 concentration exceeding 0.1% of initial procaspase-8 level. |

[4] The simulation time was 2880 min (2 days) in all predictions.
[5] The apoptotic threshold is the concentration of anti-CD95 after which cPARP amount exceeds 10% of the initial PARP level.

**Table 9S (continuation)**

<table>
<tr><th>№</th><th>Behavior of the model by Bentele et al.</th><th>Behavior of the reduced model</th><th>The composite model behavior</th><th>Experimental observations by Bentele et al. and predictions of the models</th></tr>
<tr><td>4</td><td>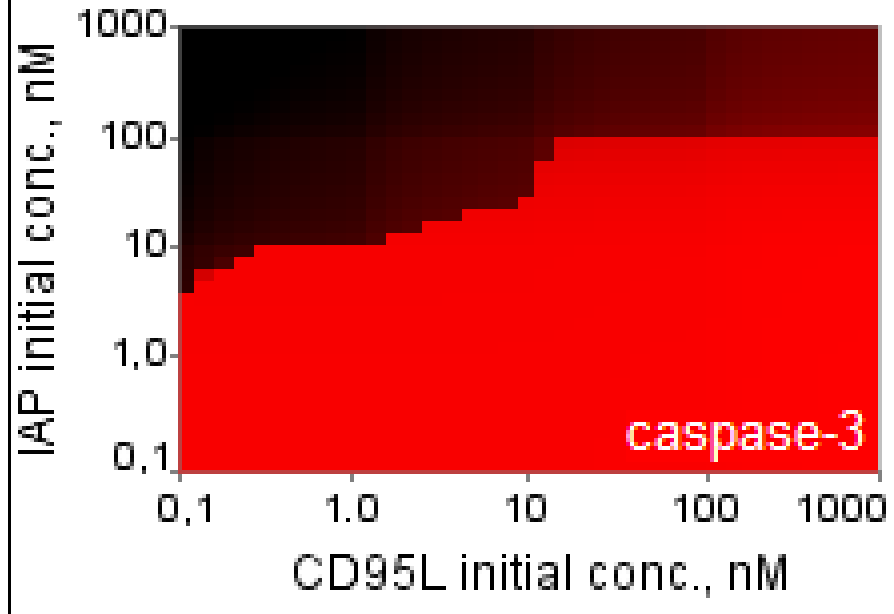
</td><td>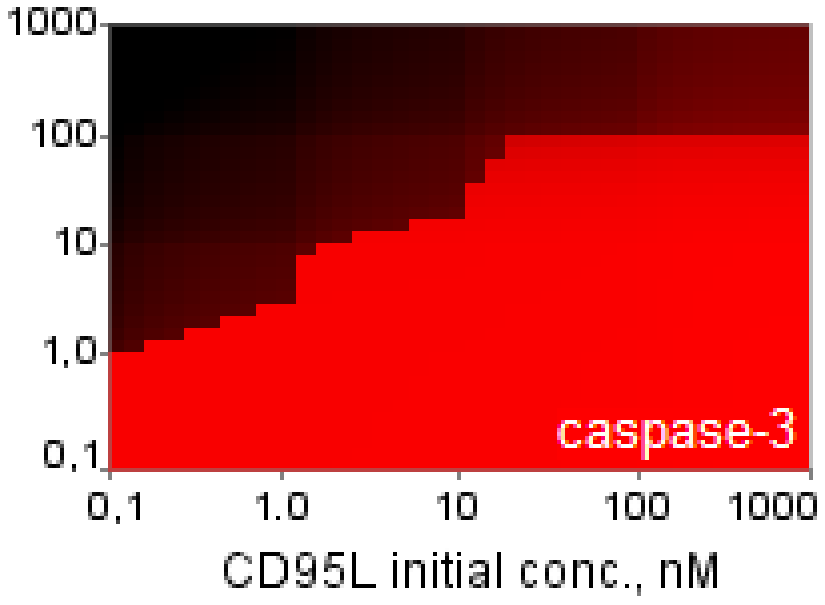
</td><td>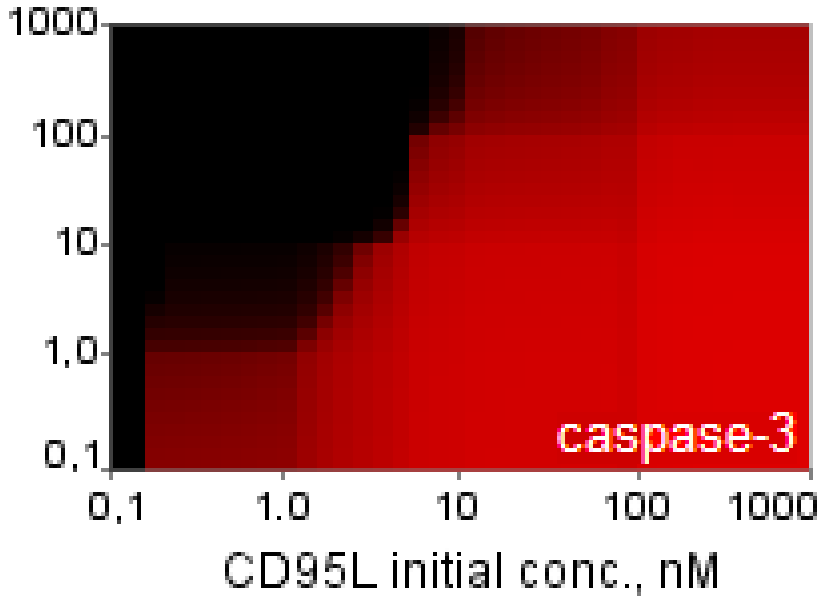
</td><td>Original and reduced models:<br>– low concentrations of IAP (less than 1 nM) result in complete cell death;<br>– high concentrations of IAP prevent significant increase of caspase-3 even for high concentrations of the ligand.<br>Composite model:<br>– low concentrations of IAP (less than 1 nM) block apoptosis for CD95L less than 0.3 nM;<br>– high concentrations of CD95L lead to cell death.<br><br>The figures show logarithmic dependence of the maximal caspases-3 concentration on initial values of IAP and CD95L.</td></tr>
</table>

**Table 10S – Parametric constraints of the models by Bentele *et al.* and Neumann *et al.***

| № | Parametric constraints of the Bentele's model[6] | Corresponding reduction steps |
|---|---|---|
| 1 | $C_{casp7} \approx a \cdot C_{casp3},\ a = const,$<br>$k_{7act} \cdot C_{casp7} \gg k_{36act} \cdot C_{casp3} > k_{36act} \cdot C_{casp6},$<br>$Km_{367act} \gg (1 - x_{aa})$ | Simplification of the equation of $x_{aa}$ production |
| 2 | $v_{br3} \approx v_{br4} \approx v_{br5} \approx v_{br6} \approx v_{br12} \approx v_{br13}$ | Elimination of quasi-stationary intermediates CD95R:CD95L, DISC:pro8, DISC:pro8$_2$ and DISC:p43/p41. Removal of slow reactions br12 and br13 |
| 3 | $C_{casp6} \approx b \cdot C_{casp3}, b = const$ | Deletion of caspase-6 and procaspase-6 |
| 4 | $C_{pro8} \gg Km_{68}$ or $v_{br6} \gg v_{br8}$ | Formation of the linear kinetic law of caspase-8 activation triggered by caspase-3 |
| 5 | $v_{br9} \approx v_{br10}$ | Elimination of the quasi-stationary species DISC:cFLIPL |
| 6 | $C_{FLIPL}(0) = C_{FLIPS}(0),$<br>reactions br9 and br11 have the same rate constant $k_{DISC_FLIP}$. | Lumping of cFLIPL and cFLIPS |
| 7 | $v_{br19}$, $v_{casp9\ degr}$ >> $v_{br20}$, $v_{br35}$, $v_{br38}$ | Deletion of slow reactions of apoptosome complex dissociation (br20), caspase-9 inhibition (br35) and casp9:IAP dissociation (br38) |
| 8 | $v_{br15}/(v_{br16} - vbr17\ ) \approx const$ | Elimination of cytochrome *C* |
| 9 | $v_{br18} \approx v_{br19}$ | Elimination of the quasi-stationary apoptosome complex |
| 10 | $v_{br22} > v_{br23}$ | Deletion of reaction br23 (Bid truncation) and caspase-2 |
| 11 | $Km_{8Bid} \gg C_{Bid}$ | Replacement of the Michaelis-Menten kinetics in br22 with the mass action kinetics |
| 12 | $v_{br27} > v_{br28},$<br>$v_{br29} \gg v_{br30}$ | Removal of the reactions of caspase-3 (br28) and caspase-7 (br30) cleavage triggered by caspase-9 |
| 13 | $v_{br27}, v_{br33}, f_{degr}(x_{aa}) \cdot C_{casp3} \gg v_{br36},$<br>$v_{br29}, v_{br34}, f_{degr}(x_{aa}) \cdot C_{casp7} \gg v_{br37}$ | Deletion of the slow reactions of casp3:IAP (br36) and casp7:IAP (br37) dissociation |
| 14 | $v_{br33}, v_{br34} \gg v_{br25}, v_{br26}$ | Taking into account the steps described above, we can get only four reactions involving IAP (br33, br34, br25 and br26). At this stage, we can remove reactions br25 and br26, as well as their members Smac and IAP:Smac. |
| 15 | $v_{br32} > v_{br31}$ | Removal of the reaction of PARP cleavage by caspase-3 |
| 16 | $Km_{78} \gg C_{pro7}$ | Replacement of the Michaelis-Menten kinetics in br29 by the mass action kinetics |
| 17 | $Km_{367act} \gg C_{PARP},$<br>$C_{casp7} \approx a \cdot C_{casp3}, a = const$ | Modification of kinetics of br32 |
| | | |
| **№** | **Parametric constraints of the Neumann's model** | **Corresponding reduction steps** |
| 1 | $k2 \cdot C_{pro8}, k3 \cdot C_{FLIPL} \gg k4 \cdot C_{FLIPS}$ | The slow reaction of the DISC binding by FLIPS (nr4) leads to production of a very small amount of DISC:FLIPS. Therefore, all reactions involving this complex (nr4, nr11-nr13) can be removed without any significant effect on the model dynamics. |
| 2 | $C_{IKK} \gg C_{p43FLIP}$ | Setting the constant value of the IKK concentration |

[6]We did not analyze the constraints on degradation rates of species.